\documentclass[10pt,aps,pre,onecolumn,superscriptaddress,floatfix,notitlepage,nofootinbib,showkeys]{revtex4-1}
\usepackage{amsmath,amssymb,amsfonts,graphicx,color,booktabs,hyperref}

\newcommand{\ea}{\emph{et al. }}

\begin{document}
\title{Epidemics of Liquidity Shortages in Interbank Markets}

\author{Giuseppe Brandi}
\affiliation{Libera Universit\`a Internazionale degli Studi Sociali (LUISS) ``Guido Carli'', 00197 Rome, Italy}
\email{gbrandi@luiss.it}

\author{Riccardo Di Clemente}
\affiliation{Massachusetts Institute of Technology, MA 02139 Cambridge, United States}

\author{Giulio Cimini}
\affiliation{IMT School for Advanced Studies, 55100 Lucca, Italy}
\affiliation{Istituto dei Sistemi Complessi (ISC)-CNR, 00185 Rome, Italy}

\date{\today}

\begin{abstract}
Financial contagion from liquidity shocks has being recently ascribed as a prominent driver of systemic risk in interbank lending markets. 
Building on standard compartment models used in epidemics, in this work we develop an EDB (Exposed-Distressed-Bankrupted) model for the dynamics of liquidity shocks reverberation between banks, 
and validate it on electronic market for interbank deposits data. We show that the interbank network was highly susceptible to liquidity contagion at the beginning of the 2007/2008 global financial crisis, 
and that the subsequent micro-prudential and liquidity hoarding policies adopted by banks increased the network resilience to systemic risk---yet with the undesired side effect of 
drying out liquidity from the market. We finally show that the individual riskiness of a bank is better captured by its network centrality than by its participation to the market, 
along with the currently debated concept of ``too interconnected to fail''. 
\end{abstract}


\maketitle

\section{Introduction}

The 2007/2008 global financial crisis has raised important concerns on the stability of the financial system, in particular regarding its complex and interconnected nature 
\cite{Lau2009,Brunnermeier2009,Battiston2015,Battiston2016,Glasserman2016}. Indeed, because of bilateral or indirect exposures between financial institutions, 
credit losses and funding shortcomings can spread through the system, with potential catastrophic consequences for the whole market 
\cite{Allen2000,Gai2011,Caccioli2014,Acemoglu2015,Greenwood2015,Gualdi2016,Barucca2016}. Within this context, much attention has been devoted to assess systemic risk in the interbank lending market, 
namely the network of financial interlinkages resulting from overnight loans between banks \cite{Rochet1996,Freixas2000,Iori2006}. Indeed, while interbank lending is crucial for banks 
to face fluctuating liquidity needs \cite{Allen2014} and properly fuel the economy \cite{Gabbi2015}, the interbank market turns out to be rather fragile, 
as intra-system cash fluctuations alone have the potential to lead to systemic defaults \cite{Smaga2016}, and exceptional liquidity shocks can lead to a complete market drought \cite{Brunnermeier2009}. 
The latter scenario in particular characterized the period after the global financial crisis: 
banks held more liquid assets in anticipation of future expected losses and liquidity needs \cite{Van2009,Gai2010,Berrospide2011}, and such a precautionary hoarding behavior caused a drastic increase 
of unsecured interbank borrowing rates for all but the shortest maturities \cite{Acharya2011}. The interbank market contraction was exacerbated even more by the massive liquidity injections from central banks, 
which instead of restoring interbank activity had the effect of replacing the market with the central banks balance sheets \cite{Heider2009,Gabrieli2014}

The stream of literature on systemic risk analysis for interbank markets has focused on two main aspects: the network-based characterization of the topological structure of the market 
\cite{Iori2006,Wells2002,Boss2004,Iori2008,Cocco2009,Finger2012,Fricke2015}, and the design of dynamics and metrics for the spread of financial distress over the network 
\cite{Lau2009,Eisenberg2001,Elsinger2006,Nier2007,Krause2012,Battiston2012,Bardoscia2015,Amini2016,Bardoscia2017,Serri2017} including 
liquidity shocks \cite{Cifuentes2005,Diamond2010,Kapadia2012,Hurd2013,Cimini2016}. In this work we add to this research field 
by proposing a modeling of liquidity-driven financial contagion in interbank networks through a compartment model similar to those commonly used in epidemiology 
\cite{Keeling2005,Pastor-Satorras2015}. In particular, by assuming that liquidity shocks propagate as an epidemic disease over the market, 
we adapt the well-known SIR ({\em Susceptible-Infected-Removed}) model \cite{Kermack1927,Chu2011,Sun2014} 
to the specific context of liquidity shocks interbank networks, and cast it as an EDB ({\em Exposed-Distressed-Bankrupted}) model. 
This is close to the work by Toivanen in the context of credit shocks \cite{Toivanen2013}.
We simulate such a dynamical model on the Italian interbank network e-MID (electronic market for interbank deposits) 
\cite{Iori2006,Iori2008,Finger2012,Fricke2015}, analyzing liquidity contagion in relation with the structure and evolution of the network, 
and defining systemic risk through the resilience of individual banks and of the overall system to liquidity shocks. 

In a nutshell, we show that the network topology of e-MID plays a fundamental and non-negligible role in the epidemics of liquidity distress, 
and that the total number of bank failures can assume very high values also when the initial shock is minimal. 
Importantly, systemic risk increases substantially just before the global financial crisis, and possibly goes back to pre-crisis values only long afterwards. 
The probability to go bankrupt turns out to be higher for banks accounting for larger shares of the interbank network; however, the number of lenders of a bank 
can statistically explain riskiness more than the total amount of money borrowed, in line with the recent paradigm shift from ``too big to fail'' to ``too interconnected to fail'' \cite{Huser2015}. 
We complement this analysis with a network-based study of e-MID, providing additional evidence that the market underwent significant changes due to the financial turmoil---with less banks 
participating to the market and executing fewer and lesser loan transactions. Liquidity hoarding had a considerable impact also on the underlying network structure: hub banks basically disappeared, 
the number of reciprocal and three-party contracts dropped, and---in spite of the reduced systemic risk---at the end the network lost its efficiency in terms money flow within banks. 

The rest of the paper is structured as follows. Section 2 recalls basic features and network characterization of the e-MID dataset, and Section 3 presents a network analysis of the data. 
The EDB contagion model is defined in Section 4, and model simulation results are reported in Section 5. Finally, Section 6 presents an econometric study of simulation outcomes, and Section 7 concludes.

\section{e-MID data}\label{sec:data}

The electronic Market for Interbank Deposits (e-MID) is a trading platform of unsecured money-market loans,\footnote{An anonymous and collateralized segment of e-MID (New MIC) 
was introduced in February 2009 with the aim of improving the liquidity distribution within the euro area \cite{Vento2010}.}
that is unique in the Euro area for being screen-based and fully electronic. e-MID covers the entire domestic overnight deposit market in Italy, but is open to both Italian and foreign banks, 
and a significant share of all liquidity deposit in the Euro area is traded through the e-MID platform \cite{Beaupain2008}. 
The dataset we have at our disposal consists of all the interbank transactions finalized on e-MID from January 1999 to September 2012. For each contract we have information about the amount exchanged, 
the date and time, the interest rate,  the IDs of the lender and of the borrower banks, and the contract maturity.\footnote{Transactions can be ON (overnight), ONL (overnight long), TN (tomorrow next), 
TNL (tomorrow next long), SN (spot next), SNL (spot next long), 1W (one week), 1WL (one week long), 2W (two weeks), 3W (three weeks), 1M (one month), 2M (two months), 3M (three months), 4M (four months), 
5M (five months), 6M (six months), 7M (seven months), 8M (eight months), 9M (nine month), 10M (ten months), 11M (eleven months), 1Y (one year). 
In this paper we use only overnight transactions, which represent the vast majority of trades (approximately 90\%).}

Because of the data structure, e-MID (and interbank markets in general) is properly represented as a directed weighted network, where interbank loans constitute the direct exposures between banks 
and allow for the propagation of financial distress in the system. Here we focus on quarterly networks obtained by aggregating transactions over three months, as this time scale is enough 
to allow the emergence of complex interaction patterns \cite{Finger2012}. To describe the network we employ the following notation. 
At quarter $t$, we have a system consisting of a set of $N_t$ banks (the {\em nodes} of the network) and of $L_t$ transactions among them---corresponding to the {\em links} connecting pairs of nodes. 
Transaction volumes are described by the $N_t\times N_t$ {\em weighted adjacency matrix} $\mathbf{W}(t)$, whose generic element $w_{ij}(t)\ge0$ ($i,j=1,\dots,N_t$) 
amounts to the overall loan that $i$ granted to $j$ (obtained by summing the single ON contracts resulting in a money flow from $i$ to $j$). Analogously, the whole pattern of connections 
is described by the $N_t\times N_t$ {\em binary adjacency matrix} $\mathbf{A}(t)$, whose generic element $a_{ij}(t)$ equals $1$ if a connection from node $i$ to node $j$ exists 
({\em i.e.}, if bank $j$ borrowed money from $i$ during $t$ so that $w_{ij}(t)>0$), and $0$ otherwise.\footnote{Note that as in e-MID banks do not lend money to themselves, it is $a_{ii}=w_{ii}=0$ $\forall i$ 
by construction.} The network built in this way is {\em directed}, meaning that in general $a_{ij}(t)\neq a_{ji}(t)$ and $w_{ij}(t)\neq w_{ji}(t)$ \cite{Iori2008}. 
In the following we will restrict our analysis to the {\em weakly connected component} of the network, defined as the largest subnetwork in which any two nodes are connected to each other 
by a finite sequence of links (regardless of their direction). This restriction is particularly useful in the context of simulating epidemic cascades, as it eliminates the possibility 
that the infection cannot spread when starting from or stopping into sink nodes or isolated communities.

\section{Analysis of network evolution}\label{sec:emp}

We now describe and analyze the network topology of e-MID, introducing a number of network quantities and showing the structural changes of the interbank market due to the global financial crisis. 
As reference period for the crisis, we use the time span when the TED spread was larger than unity (2007Q3 -- 2009Q1), which is indicated in the plots with a shaded area. 
Also, when displaying temporal trends of empirical quantities, we employ a 5-point simple symmetric moving average to smooth out short-term fluctuations and highlight longer-term trends.

\subsection{Degree and Strength}

\begin{figure}[h!]
\begin{center}
\includegraphics[width=0.495\textwidth]{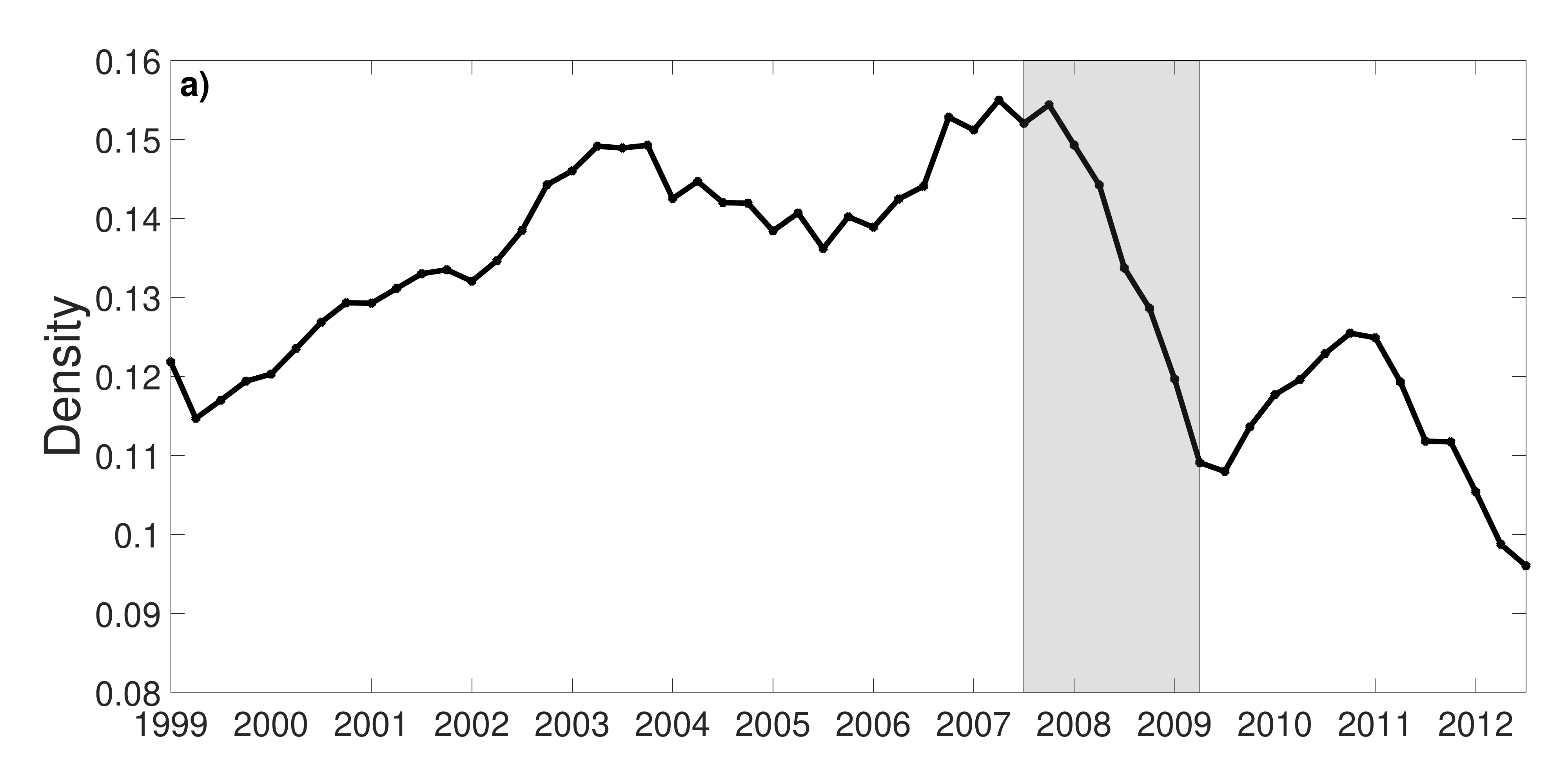}
\includegraphics[width=0.495\textwidth]{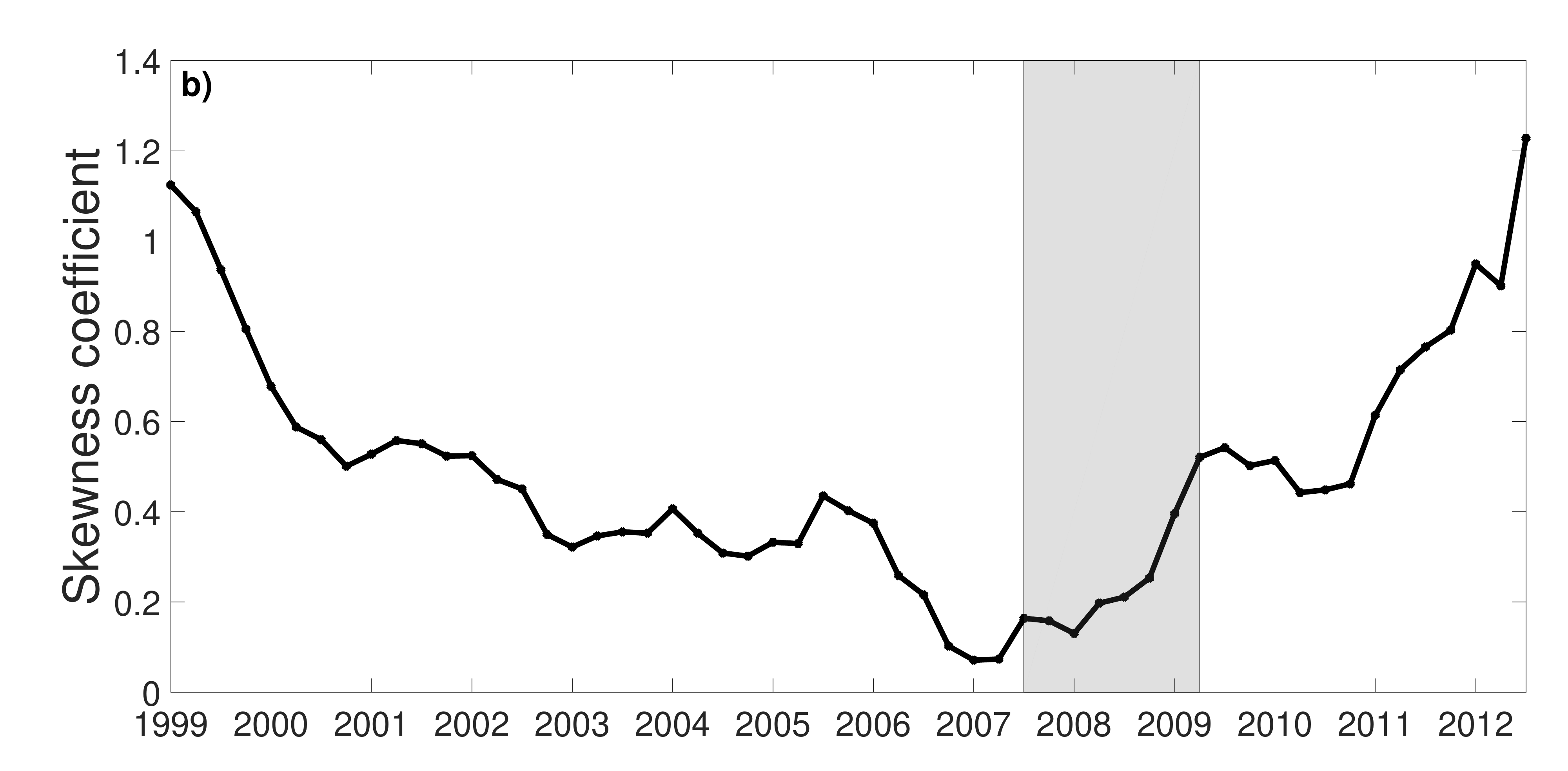}
\caption{\label{fig:1}\textbf{a) Network density in e-MID over time.} $D_t$ features a mild growth until the crisis and then falls, as several banks diminished trading on e-MID from then on. 
\textbf{b) Skewness coefficient of the degree distribution in e-MID over time.} $P(k)$ is generally positively skewed, but the distribution becomes rather symmetric just before and during the crisis.}
\end{center}
\end{figure}

For each bank $i$ at quarter $t$, its {\em in-degree} $k_{i}^{I}(t)=\sum_{j\in N_t}a_{ji}(t)$ is defined as the total number of incoming connections 
(that is, the number of banks that lent money to $i$ during $t$). Similarly, its {\em out-degree} $k_{i}^{O}(t)=\sum_{j\in N_t}a_{ij}(t)$ is the total number of outgoing links 
(that is, the number of banks that borrowed money from $i$ during $t$). The {\em reciprocal degree} $k_{i}^{\leftrightarrow}(t)=\sum_{j\in N_t}a_{ij}(t)a_{ji}(t)$ quantifies instead 
the number of banks that are at the same time borrowers and lenders of $i$. 
The overall number of trading partners is named {\em degree}, and is expressed as $k_{i}(t)=\sum_{j\in N_t}[a_{ij}(t)+a_{ji}(t)-a_{ij}(t)a_{ji}(t)]\equiv k_{i}^{O}(t)+k_{i}^{I}(t)-k_{i}^{\leftrightarrow}(t)$. 
Considering all banks, we can count the total number of links (that is, the number of bank pairs that finalized at least a transaction) during quarter $t$ as $L_t=\sum_{ij}a_{ij}(t)$, 
and define the overall link {\em density} of the network as the number of realized links divided by the maximum number of possible links ({\em i.e.}, that of a fully connected network): $D_t=L_t/[N_t(N_t-1)]$. 
Figure \ref{fig:1}a shows that $D_t$ remains quite stable until the onset of the global financial crisis, and drops significantly during it as banks started to hoard liquidity 
and reduce the number of granted loans.

Concerning the distribution of banks connectivities, e-MID is known to be characterized by a highly-skewed and long-tailed degree distribution $P(k)$, with the presence of a small number of ``hub'' banks 
with many connections and of a large number of banks with very few links \cite{Iori2008}. Figure \ref{fig:1}b shows the skewness coefficient of $P(k)$, 
which is indeed positive but features a decreasing trend before the onset of the crisis, and an increasing trend afterwards. The skewness coefficient becomes very small 
just before and during the crisis, meaning that $P(k)$ loses its asymmetry during such period: degrees are distributed uniformly, and there are no hub banks in the network. 
In such a scenario, resulting from large banks drastically reducing their market participation, the network resilience to random failures of nodes drops \cite{Iori2008}.

\begin{figure}[h!]
\begin{center}
\includegraphics[width=0.495\textwidth]{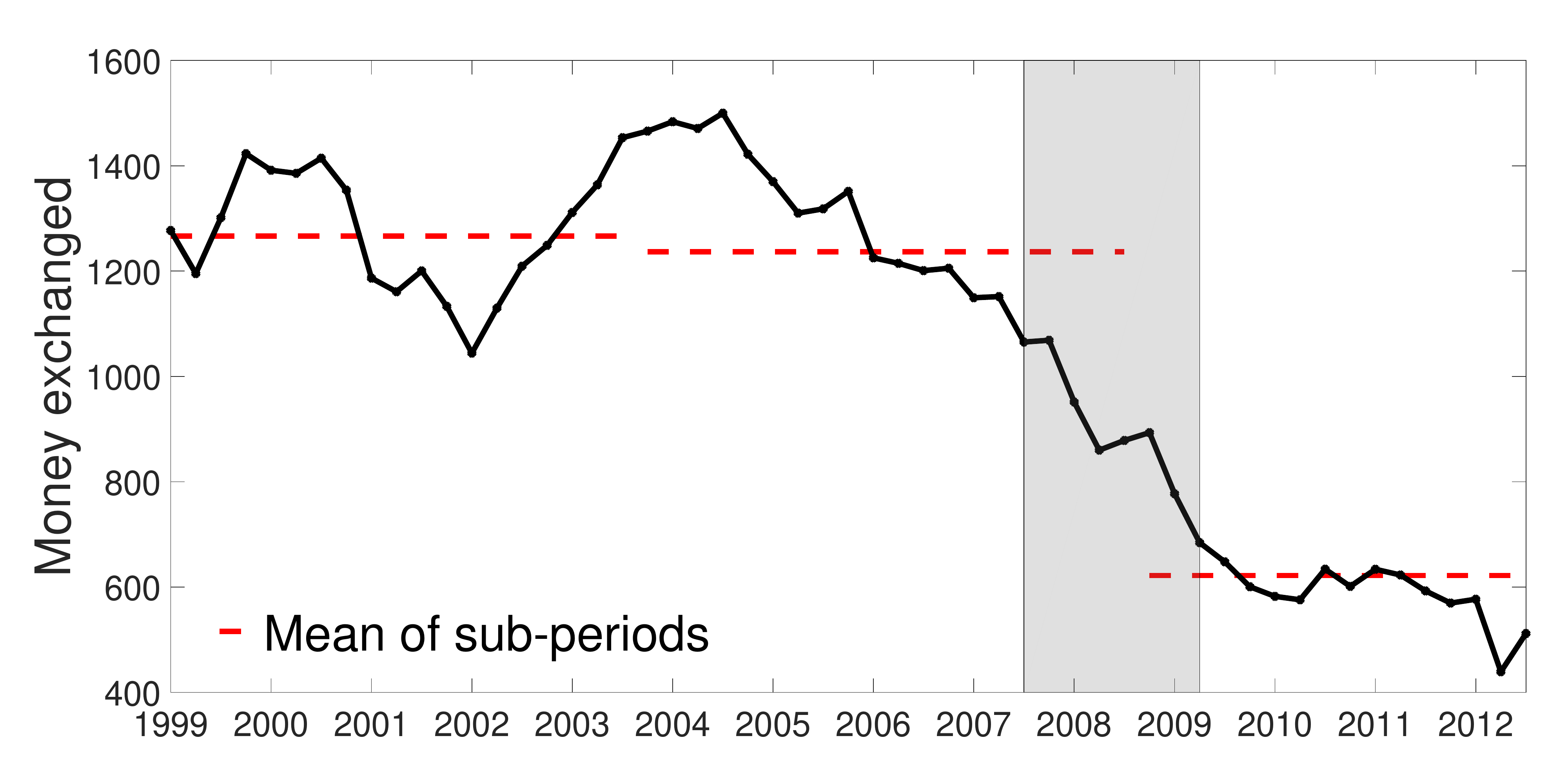}
\caption{\label{fig:2} \textbf{Average volume of money exchanged (in billions of euros) per bank over time.} Banks' participation to the market is rather stable until the crisis, and then drops consistently. 
According to a two-sample Kolmogorov-Smirnov test (at 5\% confidence level), the hypothesis that the average values before and after the crisis are derived from the same distribution 
is strongly rejected. On the contrary, when subdividing the pre-crisis period in two sub-periods, the hypothesis cannot be rejected: the trend is stable before the crisis.} 
\end{center}
\end{figure}

Moving to the weighted structure of the network, we characterize each bank $i$ by its {\em in-strength} $s_{i}^{I}(t)=\sum_{j\in N_t}w_{ji}(t)$ (the total amount of money borrowed from other banks), 
{\em out-strength} $s_{i}^{O}(t)=\sum_{j\in N_t}w_{ij}(t)$ (the total amount of money lent to other banks), and overall {\em strength} 
$s_{i}(t)=\sum_{j\in N_t}w_{ij}(t)+w_{ji}(t)\equiv s_{i}^{O}(t)+s_{i}^{I}(t)$ (the total amount of money exchanged by that bank). 
Similarly to the number of links, the total volume of trades at quarter $t$ is defined as $V_{t}=\sum_{ij}w_{ij}(t)$. 
Figure \ref{fig:2} shows that the average volume of money exchanged per bank ($V_t/N_t$) drops during the crisis: again, during turmoils banks hand out less loans 
as worries of counterparty creditworthiness prevail over the increasing possible gains due to interest rate bursts \cite{Lyons2009,De2014}. Average traded volumes remain low also after the crisis, 
supposedly because of difficulties in re-building trust between banks.

\subsection{Core-periphery structure}

\begin{figure}[h!]
\begin{center}
\includegraphics[width=0.495\textwidth]{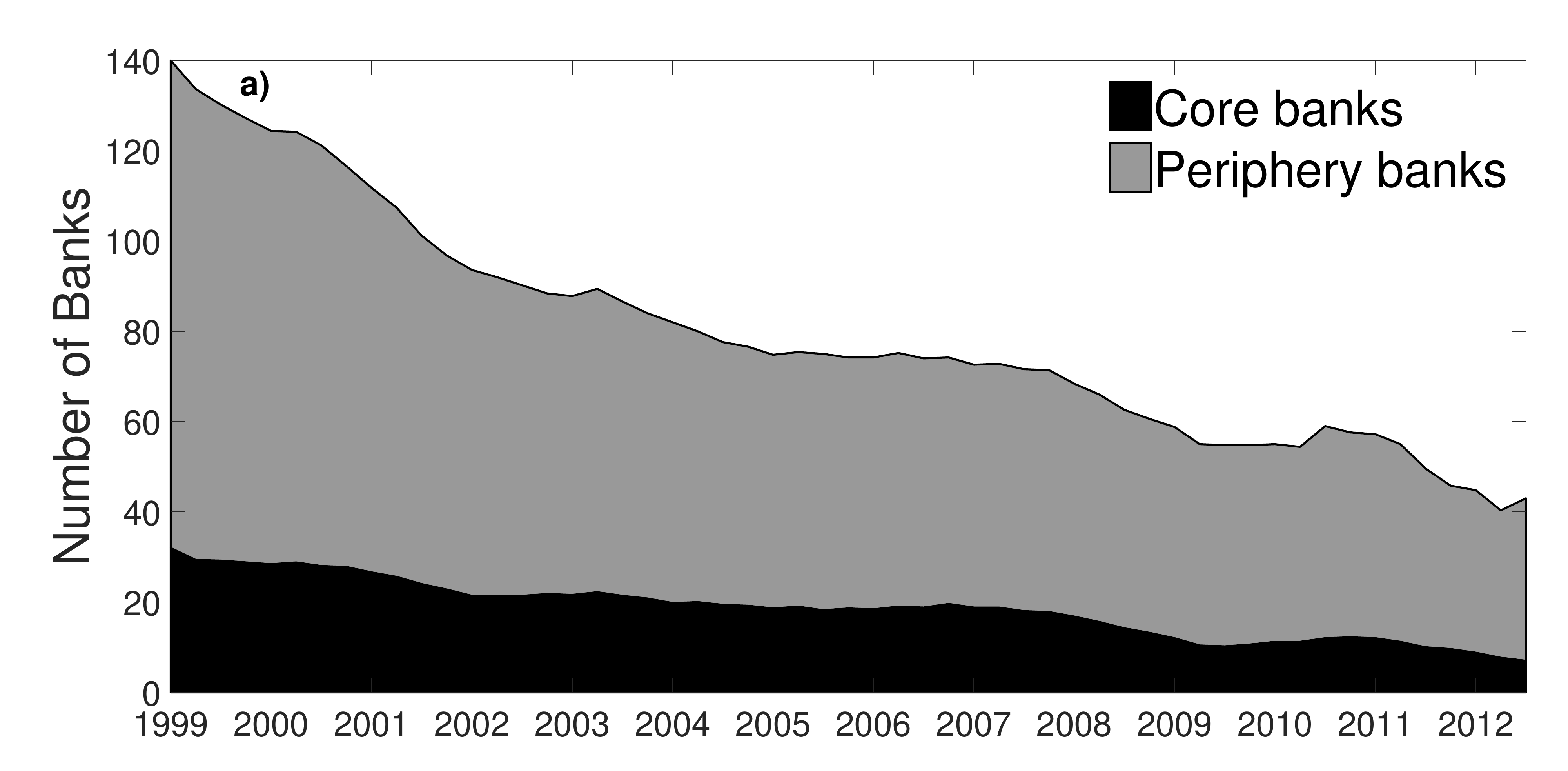}\\
\includegraphics[width=0.495\textwidth]{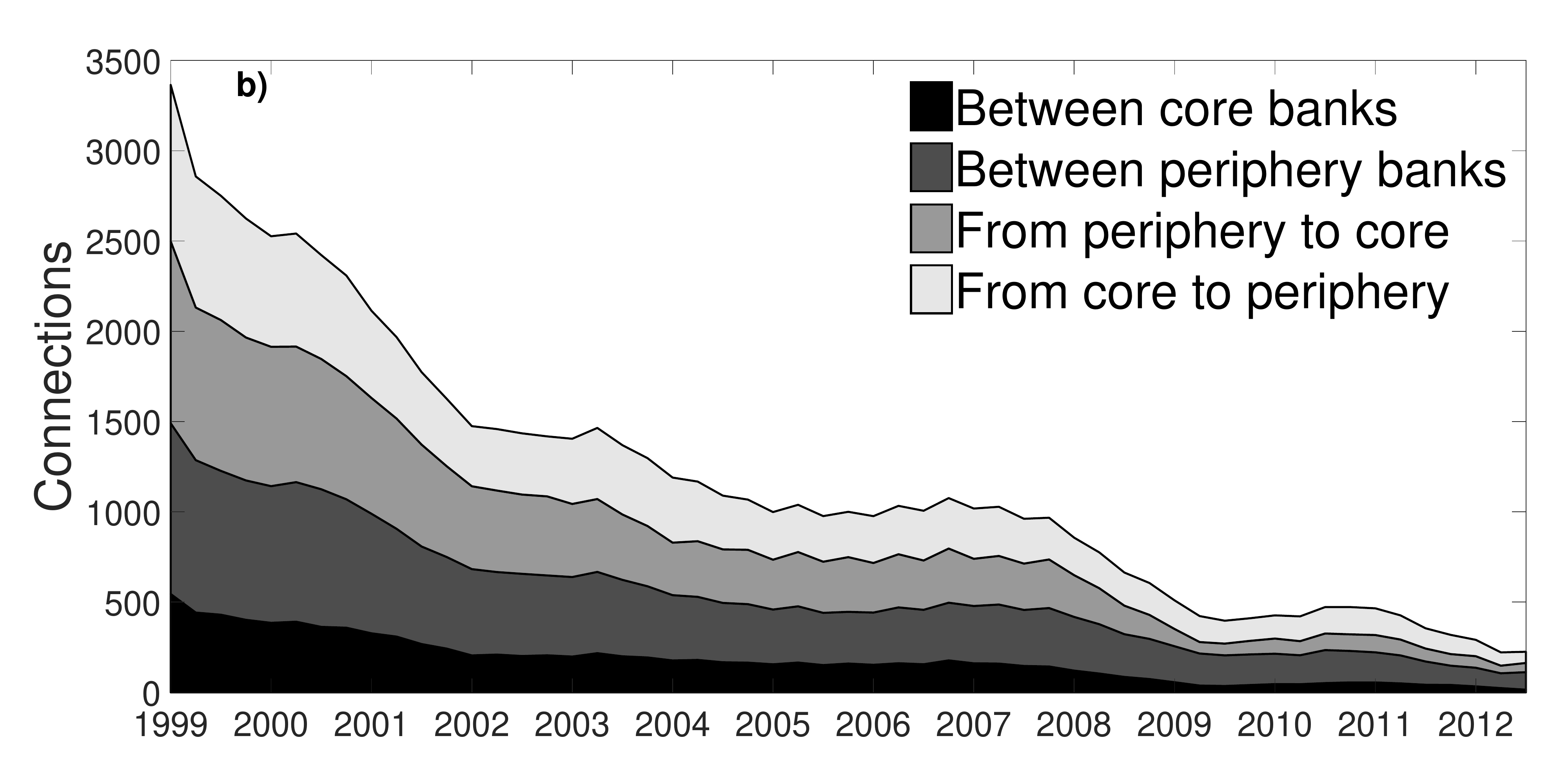}
\includegraphics[width=0.495\textwidth]{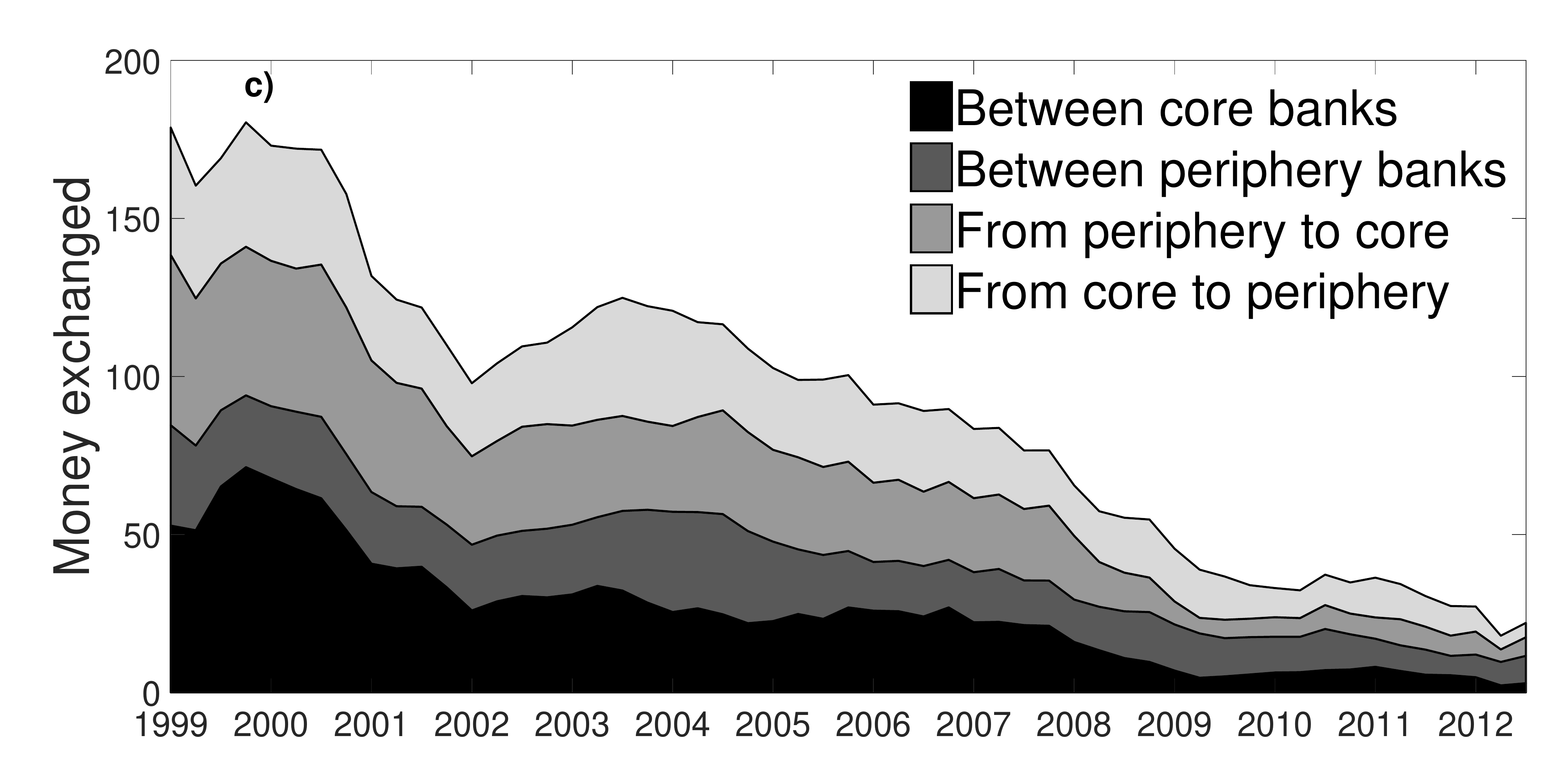}
\caption{\label{fig:3}\textbf{Core-periphery network analysis over time.} Stacked histograms for: a) Number of active banks $N_t$; b) Number of links between active banks $L_t$; 
c) Volume of money exchanged $V_t$ (in billions of euros).}
\end{center}
\end{figure}

At the quarterly aggregation level, e-MID is characterized by a {\em core-periphery structure}: hub banks (the {\em core}) are highly connected among themselves, whereas, the other leaf banks 
(the {\em periphery}) are connected mostly to the core, and very few links exist within the periphery \cite{Fricke2015}.\footnote{At aggregation levels shorter than one week, the network is instead 
better characterized by a bipartite (or {\em bow-tie}) structure, with some banks that preferentially lend and other banks that preferentially borrow \cite{Barucca2015}.} 
Here we resort to the method in \cite{Craig2014} to identify core and periphery banks, and employ the obtained partition in the analysis of overall network properties presented in Figure \ref{fig:3}. 
In particular, Figure \ref{fig:3}a shows the decreasing trend of the number of {\em active banks} $N_t$, namely the number of banks that appear at least in one e-MID transaction during quarter $t$. 
Figures \ref{fig:3}b and \ref{fig:3}c respectively show that also the number of connections $L_t$ and the volume exchanged $V_t$ decrease over time, 
whatever these quantities are computed within the core, within the periphery, and between core and periphery. 
Note that the decrease of $L_t$ and $V_t$ is only partially explained by the fewer number of active banks $N_t$ (in fact, $V_t/N_t$ is almost constant before the crisis, see Figure \ref{fig:2}). 
In this respect, while precautionary motives can be responsible for having less active banks since the onset of the crisis, mergers and acquisitions could have played a significant role beforehand. 

\subsection{Reciprocity, clustering and efficiency}

\begin{figure}[h!]
\begin{center}
\includegraphics[width=0.5\textwidth]{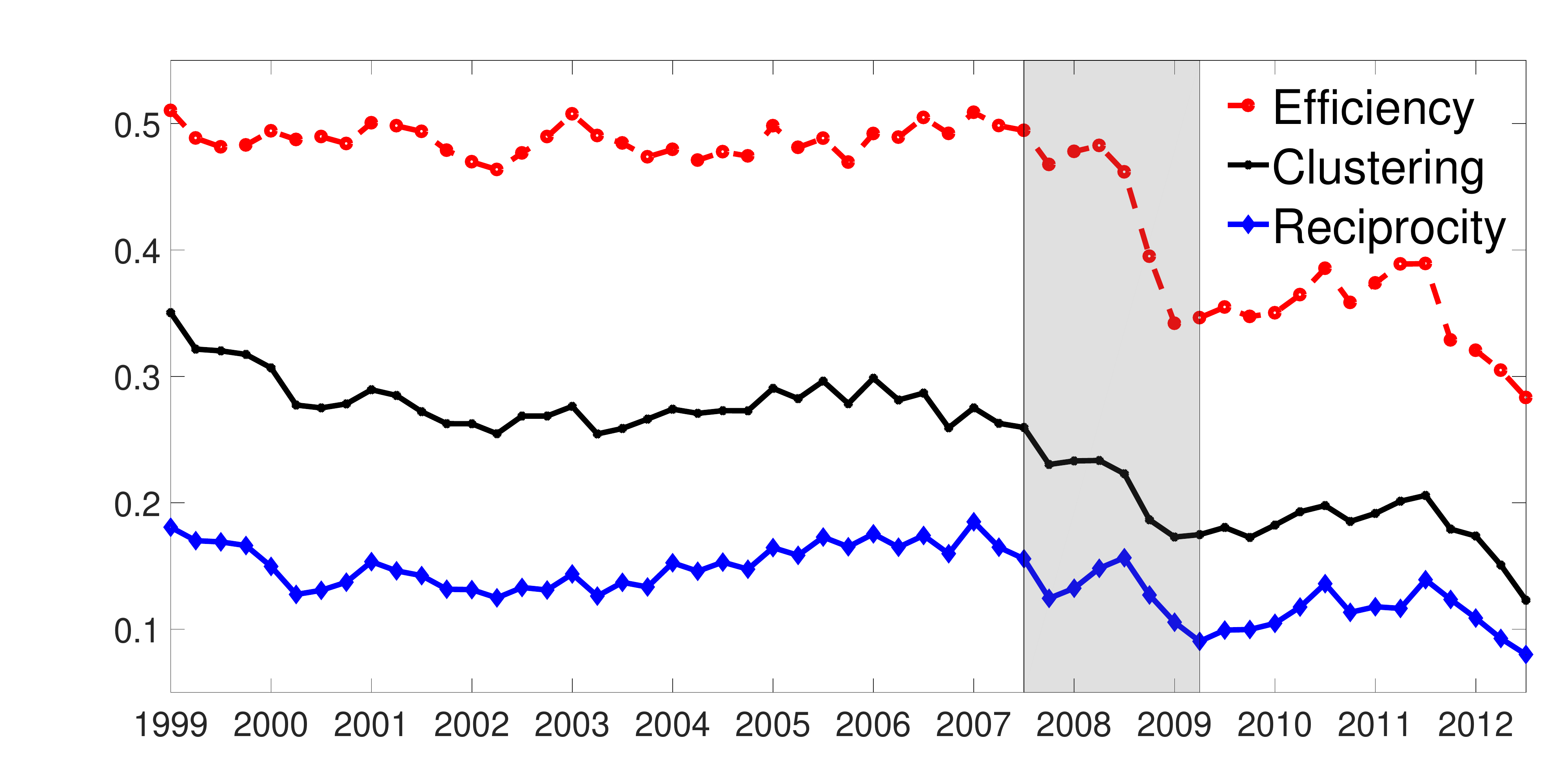}
\caption{\label{fig:4} \textbf{Reciprocity, clustering and efficiency coefficients over time.} The drop of efficiency at the global financial crisis is particularly evident, 
and basically marks the interbank market freeze \cite{Gabrieli2014}.}
\end{center}
\end{figure}

We now turn our attention to network metrics related to the local structure of connections, and in particular to preferential lending between banks. 
{\em Reciprocity} is defined as the the number of bidirectional links over the total number of links in the network: $r_t=L_t^{\leftrightarrow}/L_t$, where $L_t^{\leftrightarrow}=\sum_{i<j}a_{ij}(t)a_{ji}(t)$. 
{\em Clustering} measures the probability that two banks have a common trading counterpart, and is defined as the average ratio between the number of link triangles to the total possible number of such triangles. 
In its simplest undirected version, the clustering coefficient is defined as $c_t=\sum_i\{[\sum_{jh}u_{ij}(t)\,u_{ih}(t)\,u_{jh}(t)]/[k_{i}^2(t)-k_{i}(t)]\}/N_t$,
where $u_{ij}(t)=a_{ij}(t)+a_{ji}(t)-a_{ij}(t)a_{ji}(t)$ \cite{Fagiolo2007}. {\em Efficiency} instead measures how well liquidity (or information in general) is exchanged and flows within the network, 
and is defined as the average of the inverses of the shortest path length between node pairs: $e_t=\sum_{ij}d_{ij}^{-1}(t)/[N_t(N_t-1)]$, 
where $d_{ij}$ is the length (in number of links) of the shortest path from node $i$ to node $j$ \cite{Latora2001}.
The temporal evolution of these quantities is reported in Figure \ref{fig:4}. Again we observe a stable behavior that changes in correspondence of the financial crisis (abruptly in the case of efficiency), 
and previous stationary values are not recovered even years later. This indicates that banks behavior due to the crisis (liquidity hoarding and distrust for counterparty creditworthiness) 
causes a structural break in the market: recovering the pre-crisis configuration may need a strong exogenous intervention and system-level modifications.

\section{Liquidity-driven contagion model}

We now present the EDB dynamical model of liquidity contagion in interbank networks. 
We start by briefly describing the standard SIR model used in epidemiology, and then proceed step by step to its adaptation for interbank lending markets. 

\subsection{The Susceptible-Infected-Removed model}

The SIR model is a deterministic compartment model introduced in \cite{Kermack1927}. It considers a fixed population of individuals, each of which can be in one of three possible states: 
susceptible ($\mathcal{S}$), infected ($\mathcal{I}$) and removed (or recovered) ($\mathcal{R}$). The model dynamics then set the transition probabilities between the compartments: 
infected individuals can spread the infection to susceptible ones with a given probability $\lambda$; on the other hand, infected individuals are removed from the system at a rate $\mu$. 
The dynamical flow of SIR model is thus $\mathcal{S}\rightarrow\mathcal{I}\rightarrow\mathcal{R}$, and the model equation for a well-mixed population are:
$$\frac{ds}{dt}=-\lambda si \qquad\qquad\qquad \frac{di}{dt}=\lambda si-\mu i \qquad\qquad\qquad \frac{dr}{dt}=\mu i $$ where $s$, $i$ and $r$ are the densities of susceptible, 
infected and removed individuals, respectively, such that $s+i+r=1$. This model can be implemented on any network structure by restricting the $\mathcal{S}\rightarrow\mathcal{I}$ 
transitions to realized links: infected individuals can only transmit the disease to their susceptible neighbors. 

The SIR model builds on several simplifying assumptions such as that of a link-independent infection rate and a node-independent removal rate, 
which can be appropriate for humans having similar immune responses but fails in other contexts where contagion channels can be of very different extent. 
A recent attempt to extend the SIR model to the more realistic scenario of weighted directed networks assumes a constant removal rate $\mu$, 
but the infection probability becomes dependent on link weights: 
the probability that the infection spreads from an infected individual $i$ to her susceptible neighbor $j$ reads $\lambda_{ij}=\lambda k_i^{O} w_{ij}/s_i^{O}$, 
meaning that outgoing links with larger weights become preferential contagion channels \cite{Chu2011}. 

\subsection{The Exposed-Distressed-Bankrupted model}

Building on the SIR model and its adaptation to weighted directed network, we now assemble our model for liquidity-driven financial contagion: the {\em Exposed-Distressed-Bankrupted} model (EDB). 
We start by briefly recalling the features of the interbank lending market, which is the context we want to describe. The market allows banks to face liquidity fluctuations 
and meet reserve requirements at the end of the trading day, as banks temporarily short on liquidity borrow money from other banks having excess liquidity. 
The majority of contracts within the market has overnight duration, meaning that they are placed, and shortly after resolved and rolled-over---as banks may need additional liquidity 
to refinance their activities. The system is thus subject to {\em roll-over risk}: banks debt that is about to mature needs to be replaced by a new debt, 
yet if because of a crisis the market offer for fundings drops, banks may fail to meet their short-term obligations and be forced to enter bankruptcy, 
or to sell their illiquid assets resulting in potentially large losses during fire sales~\cite{Brunnermeier2009}. These spillovers create the incentive to hoard liquidity \cite{Berrospide2011}, 
which can in turn induce another wave of liquidity shocks, activating a spiral that can lead the interbank market to freeze completely~\cite{Acharya2011,Gabrieli2014}. 

Once the liquidity risk dynamics of the market has been set down, we can formulate the analytics of the EDB model. We basically assume that liquidity losses propagate as an epidemic disease over the market: 
funding losses experienced by a bank imply a decreasing capability to lend money to the market (even if no default has occurred), which in turn may cause liquidity shortages for other banks 
unable to borrow all the liquidity they need. A simple way to translate these concepts into a dynamical system is the following. 
Similarly to SIR, each bank can be in one of three states: exposed ($\mathcal{E}$), distressed ($\mathcal{D}$) and bankrupted ($\mathcal{B}$). 
Exposed banks are healthy in the sense that they get full liquidity supply from the market, but are subject to financial contagion from both distressed and defaulted banks. 
If contagion occurs by losses of liquidity provision, an exposed bank becomes distressed. Finally, when liquidity losses become overwhelming, a distressed bank eventually fails 
and switches to the bankrupted state. 

The ``infection'' probability, {\em i.e.}, the probability that a distressed or defaulted bank $i$ transmits a liquidity shock to an exposed bank $j$, depends on the transaction volume 
between the two institutions. In particular, we assume that the higher the loan $i$ granted to $j$ ($w_{ij}$), the higher the probability that $i$ (which is short of liquidity) cuts the funding, 
resulting in a liquidity loss for $j$. In general we assume:
\begin{equation}
 \lambda_{ij}=\varphi\left(\frac{w_{ij}}{s_i^{O}}\right)
 \label{eq.infect}
\end{equation}
where $\varphi(\cdot)$ is an appropriately chosen function defined in the interval $[0,1]$ that obeys $\varphi(0)=0$ and $\varphi(1)=1$.\footnote{For a linear $\varphi(\cdot)$, 
eq. (\ref{eq.infect}) is analogous to the expression used in \cite{Chu2011}---apart from a pre-factor and the dependence on the out-degree of the source node.} 
The role of the normalization factor $s_i^{O}=\sum_jw_{ij}$ is to have $\lambda_{ij}\equiv1$ in the case of a distressed or defaulted bank $i$ lending money only to one exposed bank $j$. 
Note that differently from the original SIR model or the one in \cite{Chu2011}, in our setting the infection propagates from both distressed and bankrupted banks.

On the other hand, the ``bankruptcy'' probability for distressed bank $i$ is assumed to be:
\begin{equation}
 \mu_i=\psi\left(\frac{\sum_{j\in N\setminus\mathcal{E}}w_{ji}}{\sum_{j}w_{ji}}\right)\equiv\psi\left(\frac{\tilde{s}_i^I}{s_i^I}\right)
 \label{eq.default}
\end{equation}
where $N\setminus\mathcal{E}$ is the set of distressed and bankrupted banks, $\tilde{s}_i^I$ is the in-strength of $i$ restricted to this set, 
and $\psi(\cdot)$ is another appropriately chosen function defined in the interval $[0,1]$ that obeys $\psi(0)=0$ and $\psi(1)=1$. 
The ratio $\tilde{s}_i^I/s_i^{I}$ is precisely the proportion of liquidity bank $i$ needs that was previously obtained from unhealthy institutions: 
a bank goes bankrupt if this proportion is too high, given the assumption that banks cannot efficiently and immediately reallocate their asset and liabilities to absorb the distress. 
Indeed, bank $i$ gets no liquidity at all if each of its lenders is in the distressed or bankrupted state, going bankrupt with probability~1. 

We stress that the EDB model builds on two important assumptions. 
The infection probability of eq. (\ref{eq.infect}) depends only the amount of the loan between the borrower and the lender (normalized to the aggregate loans of the lender), 
without considering any other features of the borrower or the lender, for instance how risky they are. 
In the same way, the bankruptcy probability eq. (\ref{eq.default}) depends only on the relative amount of funding lost by the bank, 
without considering any other features of the bank such as its liquid asset holdings. 
These assumptions are rooted on the fact that the model uses as input only the structure of the interbank market, without considering the detailed balance sheet composition of each bank. 
Such a simplification allows to run the model when the latter information is missing, yet leaves room for improvements. 
Finally note that the model does not contemplate any recovery mechanism: once a bank becomes distressed, it will eventually get bankrupted. 
As such, the model dynamic is meant to run on short time scales (comparable to loan duration which is typically ON), 
for which it is unrealistic to expect that banks can adjust their financial positions and thus recover.

While we leave the discussion on how to set the functions $\varphi(\cdot)$ and $\psi(\cdot)$ to the next section,\footnote{The model can be further extended by adding a stochastic noise 
$\epsilon$ to the infection and default probabilities: $\varphi(x)\rightarrow\varphi(x)+\epsilon$ and $\psi(x)\rightarrow\psi(x)+\epsilon$. 
Such a noise simulates instantaneous exogenous shocks, generated for instance by private depositors withdraws. The upper bound of the shock can be calibrated 
as the maximum amount of withdraw allowed by a bank under distress. For the sake of simplicity, in this work we do not consider this possibility.} 
we conclude by discussing related SIR-like models in the context of economic and financial networks. 
Garas \ea \cite{Garas2010} use a variable infection probability to model the spreading of a crisis in the world economy. 
Toivanen \cite{Toivanen2013} and later Philippas \ea \cite{Philippas2015} model interbank contagion from the viewpoint of {\em counterparty risk}, 
due to losses originating from banks failing to meet contractual obligations. As such, this kind of contagion travels in the opposite direction 
of contagion due to roll-over risk which we consider in our study \cite{Lau2009,Cimini2016}. Indeed, Toivanen \cite{Toivanen2013} models contagion from bank $i$ to bank $j$ 
using an infection strength $\kappa_{ij}\sim w_{ji}/s_j^{O}$ (namely, the share of the loan to $i$ in the portfolio of bank $j$), 
which must be higher than a parameter reflecting the financial position of bank $j$ for contagion to occur. 
Overall, our liquidity-driven approach represents a natural complement to the credit-driven model by Toivanen \cite{Toivanen2013}.

\subsection{Simulation setting}

The numerical program for the EDB model is designed as follows. 
Initially, all banks are healthy and thus belong to the exposed state $\mathcal{E}$. Then, an exogenous liquidity shock hits a set of randomly chosen banks, 
which are thus put into the distress state $\mathcal{D}$. The dynamics then proceeds in discrete time steps, in each of which:
\begin{itemize}
\item Distressed and bankrupted banks can infect their exposed debtor banks, with probability given by eq.~(\ref{eq.infect}). 
If infected, an exposed bank switches to the distressed state (and can thus propagate liquidity shocks in the next step).
\item After the propagation of the infection, distressed banks can go bankrupt at a rate given by eq.~(\ref{eq.default}).
\end{itemize}
The simulation stops when there are no distressed banks left in the market, or if a maximum number of steps is reached. 
The stop condition of no distressed banks left in the market means that, from then on, no more defaults can occur---unless an exposed bank becomes distressed first. 
The upper limit of time steps is instead set as the model is designed to run on short time scales.\footnote{We arbitrarily set this upper limit to 100 steps, 
noting that about 90\% of simulations stop before reaching it.}

\begin{figure}[h!]
\begin{center}
\includegraphics[width=0.495\textwidth]{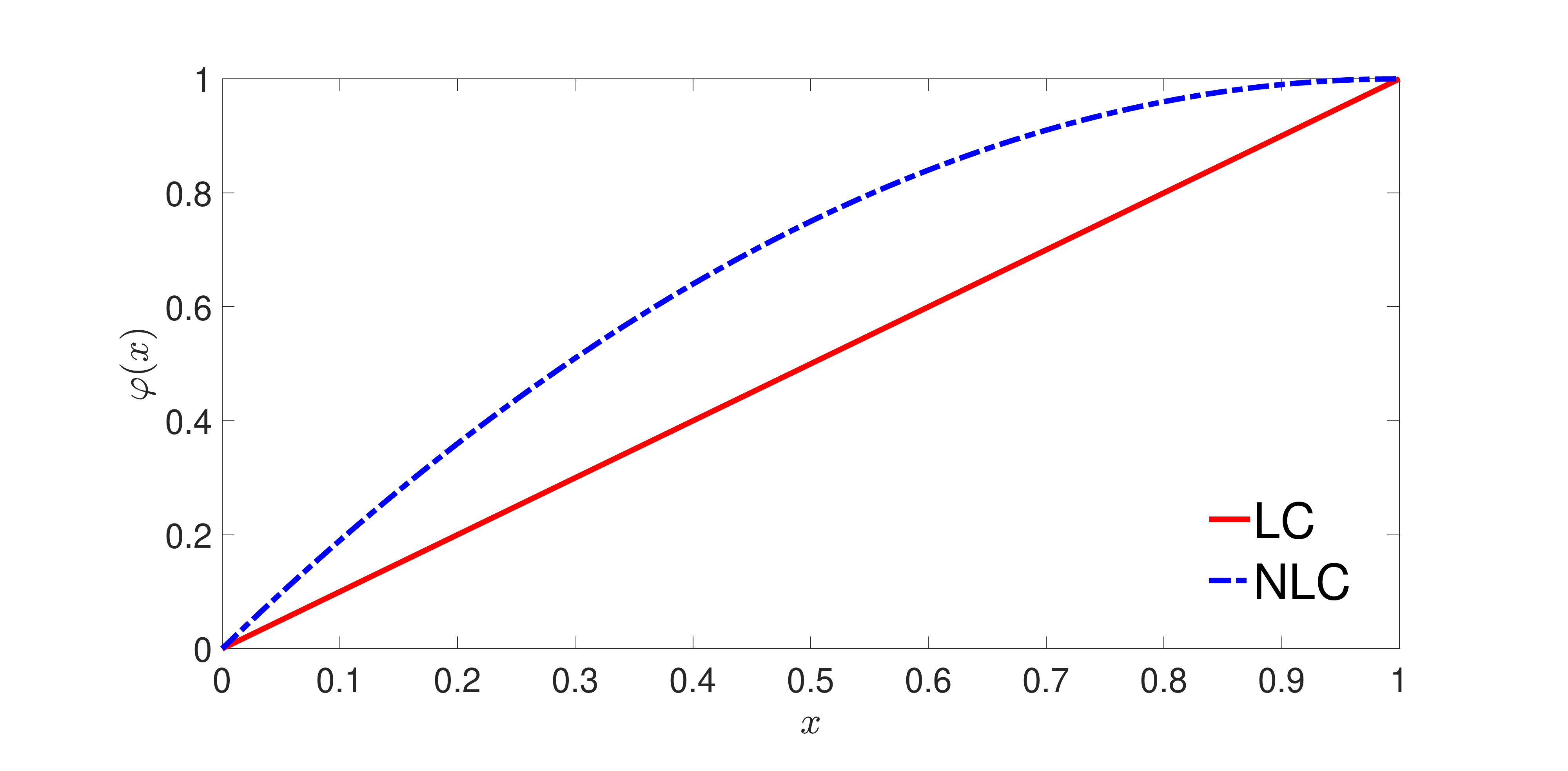}
\includegraphics[width=0.495\textwidth]{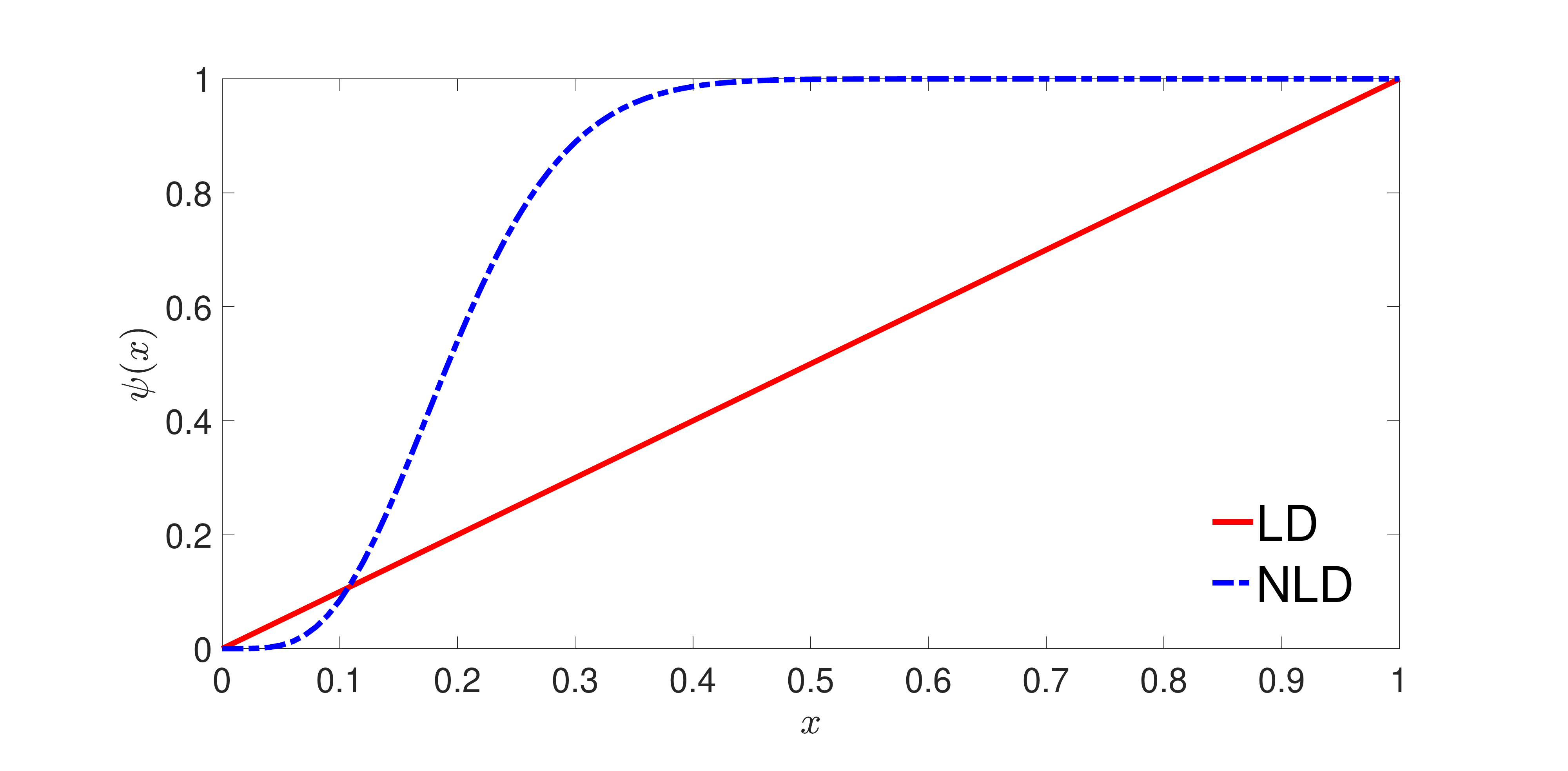}
\caption{\label{fig:X} \textbf{Economic scenarios modeled by the functions $\varphi(\cdot)$ and $\psi(\cdot)$.} 
For NLD we use a regularized incomplete beta function $\psi(x)=I_x(5,20)$, whereas, for NLC we pose $\varphi(x)=I_x(1,2)$.}
\end{center}
\end{figure}

For our analysis, we implement three different economic scenarios by changing the shape of the functions $\varphi(\cdot)$ and $\psi(\cdot)$ (see Figure \ref{fig:X}):
\begin{enumerate}
\item LC-LD (linear contagion -- linear default). The infection probability is linearly proportional to the exposure between banks, and the default probability 
is linearly proportional to the ratio between the faced liquidity shortage and the overall financial needs of a bank, resulting into linear expressions $\varphi(x)=\psi(x)=x$. 
\item LC-NLD (linear contagion -- nonlinear default). The infection probability is linear $\varphi(x)=x$, but the default probability is a nonlinear function characterized by a threshold $\gamma$, 
so that $\psi(x)\propto x$ for $x<\gamma$ and $\psi(x)\simeq 1$ for $x\ge\gamma$.
The latter choice reflects a situation in which banks with little liquidity shortages have easy access to funding 
(from external channels such as central banks injecting liquidity in the market) and are thus less likely to go bankrupt, whereas, the default probability jumps 
when liquidity shortages surpass a threshold, as access to external funding is denied for exceeding regulatory constraints. 
\item NLC-NLD (nonlinear contagion -- nonlinear default). The infection probability grows more than linearly with the exposure, mimicking a liquidity hoarding tendency for distressed banks 
which puts more liquidity pressure on their counterparties, whereas, the default probability is kept nonlinear.
\end{enumerate}

As input data for simulations, we use quarterly networks as described in section \ref{sec:data}. 
Note that the use of quarterly networks is indeed consistent with a model dynamics run on short time scales. 
This is because the EDB model is only based on connection weights relative to banks exposure. 
In a typical daily network a bank establishes a few contracts (so that there are a few channels of financial contagion), 
but each of these contract represents a significant share of its total daily exposure (and is thus very likely to drive liquidity contagion). 
In a quarterly network instead a bank has many more connections (so that there are several channels of financial contagion), 
though each of them weights much less in its total quarterly exposure (and is thus much less likely to drive liquidity contagion). 
Overall, for our model using a quarterly network basically means using an average daily network in the whole quarter, 
getting rid of the sparsity issue for daily networks which can lead to extremely noisy results. 
More importantly, connections in the average daily networks should more accurately represent preferential, long-lasting relationships between banks 
(which are the actual drivers of liquidity contagion in our model), whereas, actual daily transactions are akin to random draws from this underlying network \cite{Finger2012}. 
Indeed, as Finger \ea argue \cite{Finger2012}, ``for the purpose of identifying [...] systemically important institutions, longer aggregation periods may be more sensible [...]. 
Moreover, from an economic viewpoint, overnight loans can be seen as longer-term loans, where the lender can decide every day whether to prolong the loan or not. 
In this way, aggregating overnight credit relationships over longer frequencies could provide a more accurate picture of the interbank network''.

\section{Simulation results}

We now present simulation results obtained as averages over 5000 independent realizations of the model.
As in section \ref{sec:emp}, we use 2007Q3 -- 2009Q1 to mark the crisis, and employ a 5-point smoothing on simulated quantities to mitigate short-term fluctuations.

\begin{figure}[h!]
\begin{center}
\includegraphics[width=0.495\textwidth]{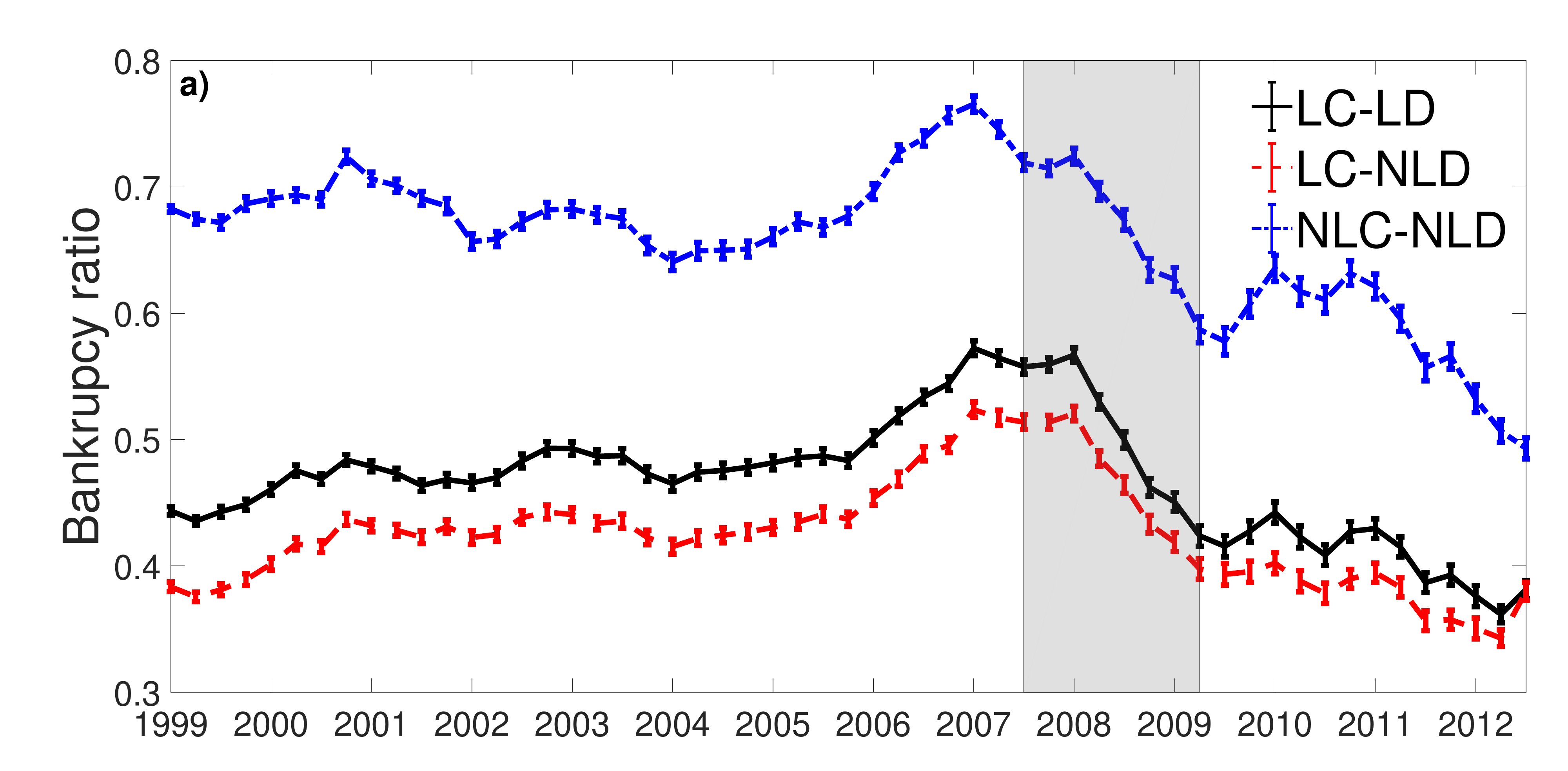}\\
\includegraphics[width=0.495\textwidth]{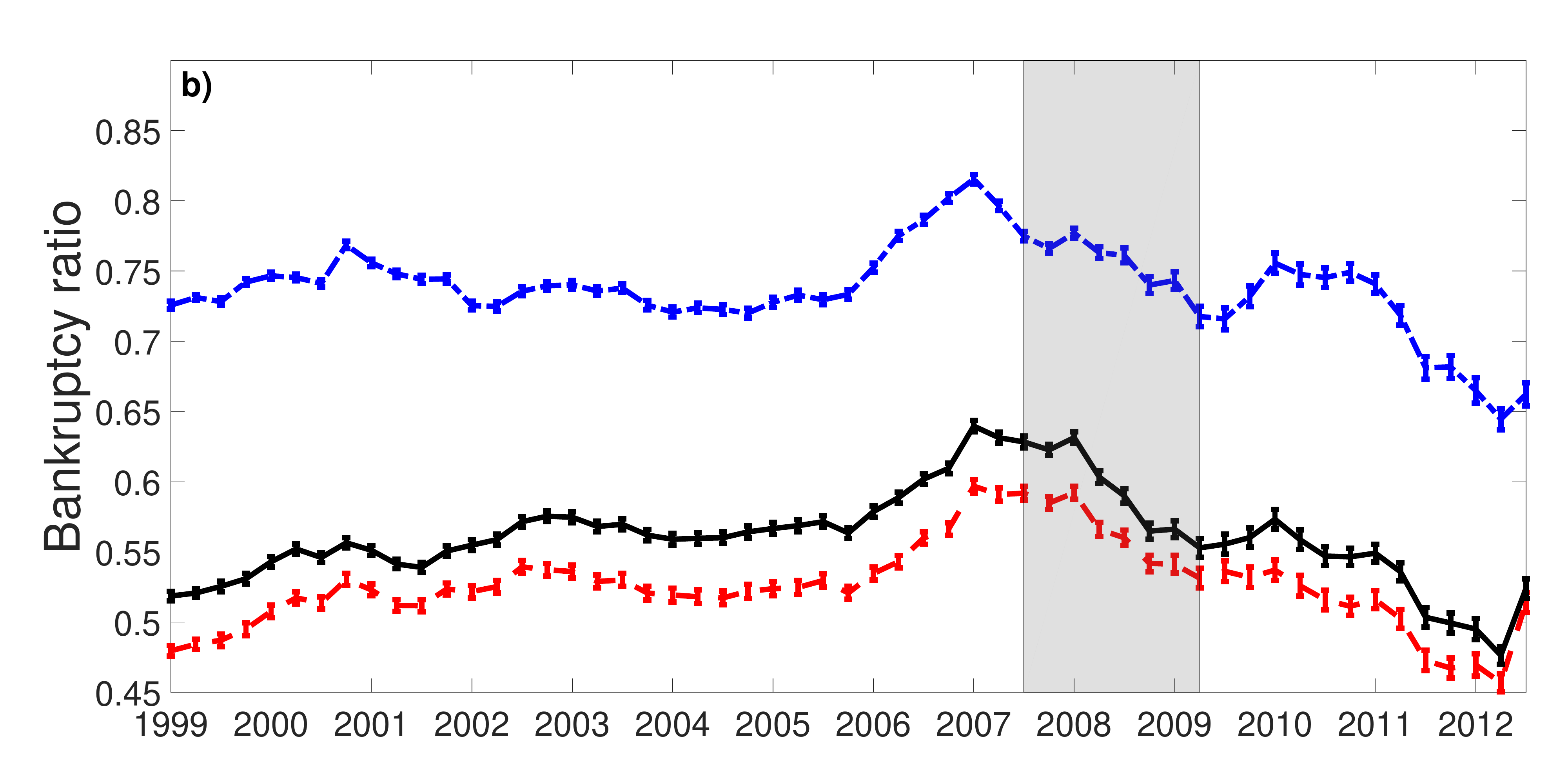}
\includegraphics[width=0.495\textwidth]{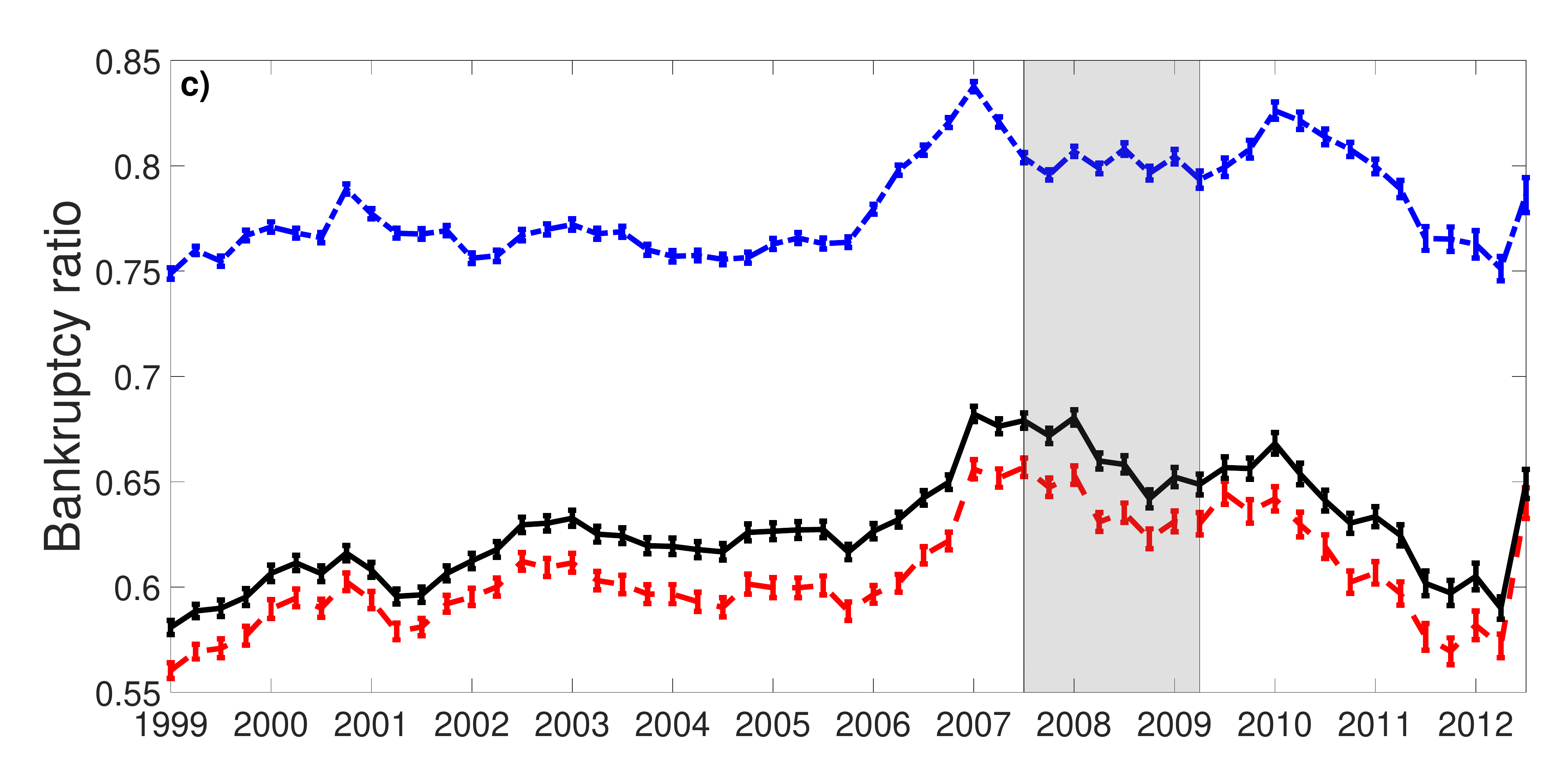}
\caption{\label{fig:5} \textbf{Average ratio of bankrupted banks at the end of simulations, for the various economic scenarios and different initial densities of distressed banks:} 
a) 1\%, b) 5\%, c) 10\%. The number of defaults reaches a maximum just before the beginning of the financial crisis.}
\end{center}
\end{figure}

We first focus on the fraction of bankrupted banks in the market, which we take as a proxy for systemic risk. 
Figure \ref{fig:5} shows simulation results for different initial densities of distressed banks and for the three economic scenarios presented above. 
The first evident observation is that the temporal trends of systemic risk for the three economies are highly correlated, and basically differ just for a vertical shift. 
This suggests that the network topology (which is the same in the three settings) plays a fundamental role in the epidemic diffusion. 
The overall level of systemic risk in the system can assume worrying high values (from about 35\% up to almost 80\% also when the initial density of distressed bank is just $1\%$), 
pointing to the importance of considering network effects when assessing the system resilience to liquidity shocks. 
Naturally, the resulting systemic risk increases for higher initial densities of distressed banks, but the relationship between these two variables is nonlinear: 
the number of bankruptcies grows only by a few percentage points (on average, 6\%, 12\% and 19\%) for multiples of the initial density (5, 10 and 20 times, respectively), 
indicating that liquidity losses in the market can saturate even for a small initial distress. In any event, the most interesting feature of Figure \ref{fig:5} is that 
the number of bankrupted banks increases substantially just before the onset of the global financial crisis. 
The subsequent drop of systemic risk during and after the crisis, which is observed for a low initial shock, can be indeed attributed to precautionary banks behavior, 
as demonstrated by the decreasing number of banks participating to the market and the even more decreasing number and volume of transactions (see Figure \ref{fig:3}). 
Yet, the reduced network density is not enough to explain the recovered resilience, and the structural changes of the market topology have been determinant in this sense. 
In order to prove this statement, we perform a statistical test by measuring systemic risk as simulated on an ensemble of null networks, {\em i.e.}, random networks with the same density, 
degree sequence and strength sequence of e-MID network snapshots after the crisis \cite{Squartini2011,Cimini2015}. Using a two-sample Kolmogorov-Smirnov test, 
the null hypothesis that the two risk measures come from independent random samples of the same distribution is rejected at the $0.1\%$ confidence level. 
Note however that the system resilience is not recovered after the crisis when the initial shock is large.

Note that an apparent contradiction to the above discussed relation between liquidity hoarding and systemic risk emerges 
by observing the remarkably higher number of defaults in the third scenario, where a sort of hoarding tendency enters in the contagion dynamics: 
the probability that banks cut their liquidity provision is higher than in the other economies, significantly enhancing the spreading of the liquidity disease. 
What happens here is that a liquidity hoarding behavior simulated on a given market configuration has the effect of increasing liquidity risk. 
However, if banks actually hoard liquidity then the realized market structure changes, and the new configuration is more resilient to liquidity shocks, 
mainly because is deprived from its functionality of liquidity provider to banks.

\begin{figure}[h!]
\begin{center}
\includegraphics[width=0.495\textwidth]{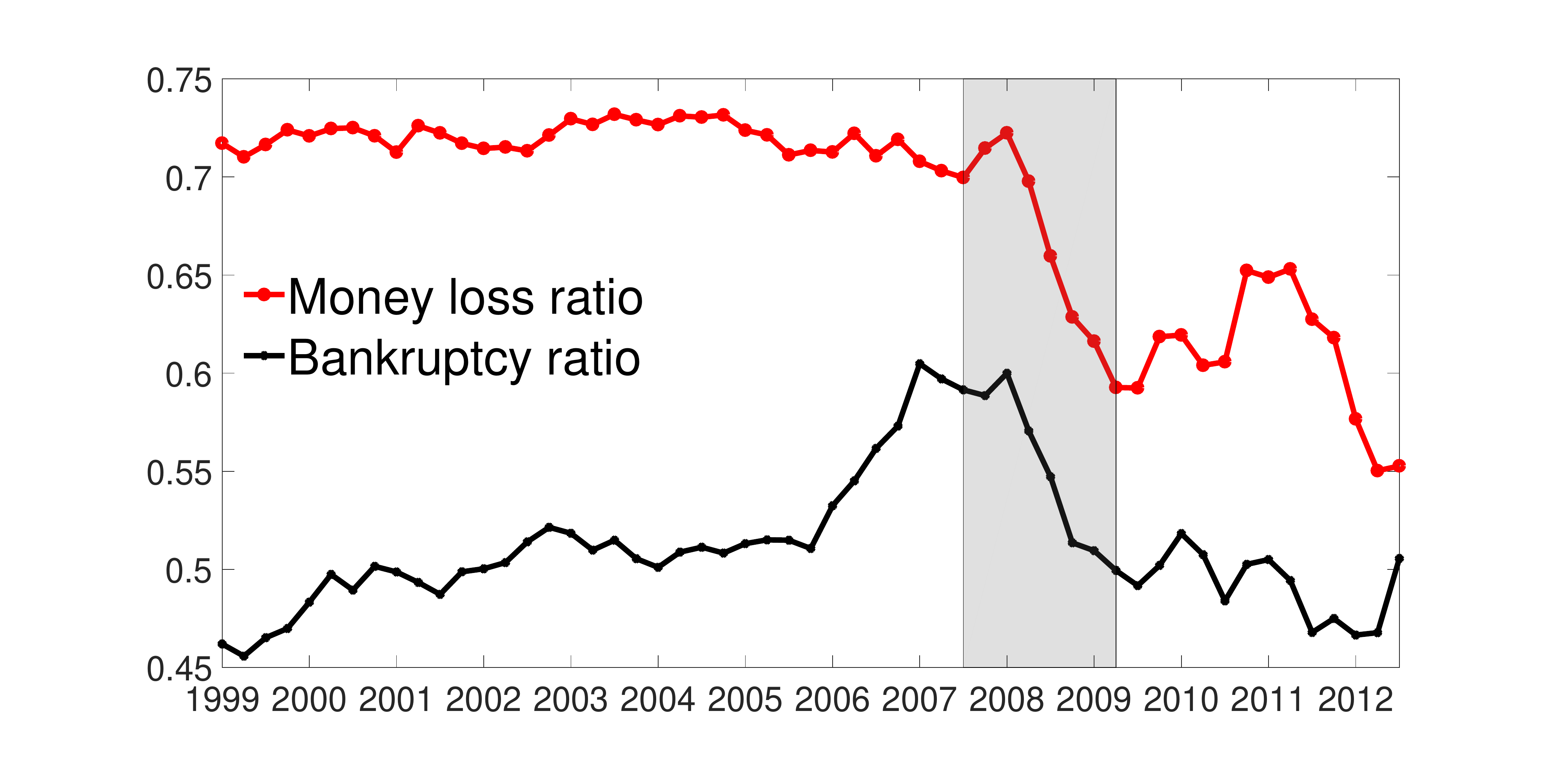}
\caption{\label{fig:6} \textbf{Relative liquidity loss compared to systemic risk given by the ratio of bankrupted banks.} 
Results refer to the first economic scenario (LC-LD) and to a density of initially distressed banks equal to 1\%.} 
\end{center}
\end{figure}

\begin{figure}[h!]
\begin{center}
\includegraphics[width=0.495\textwidth]{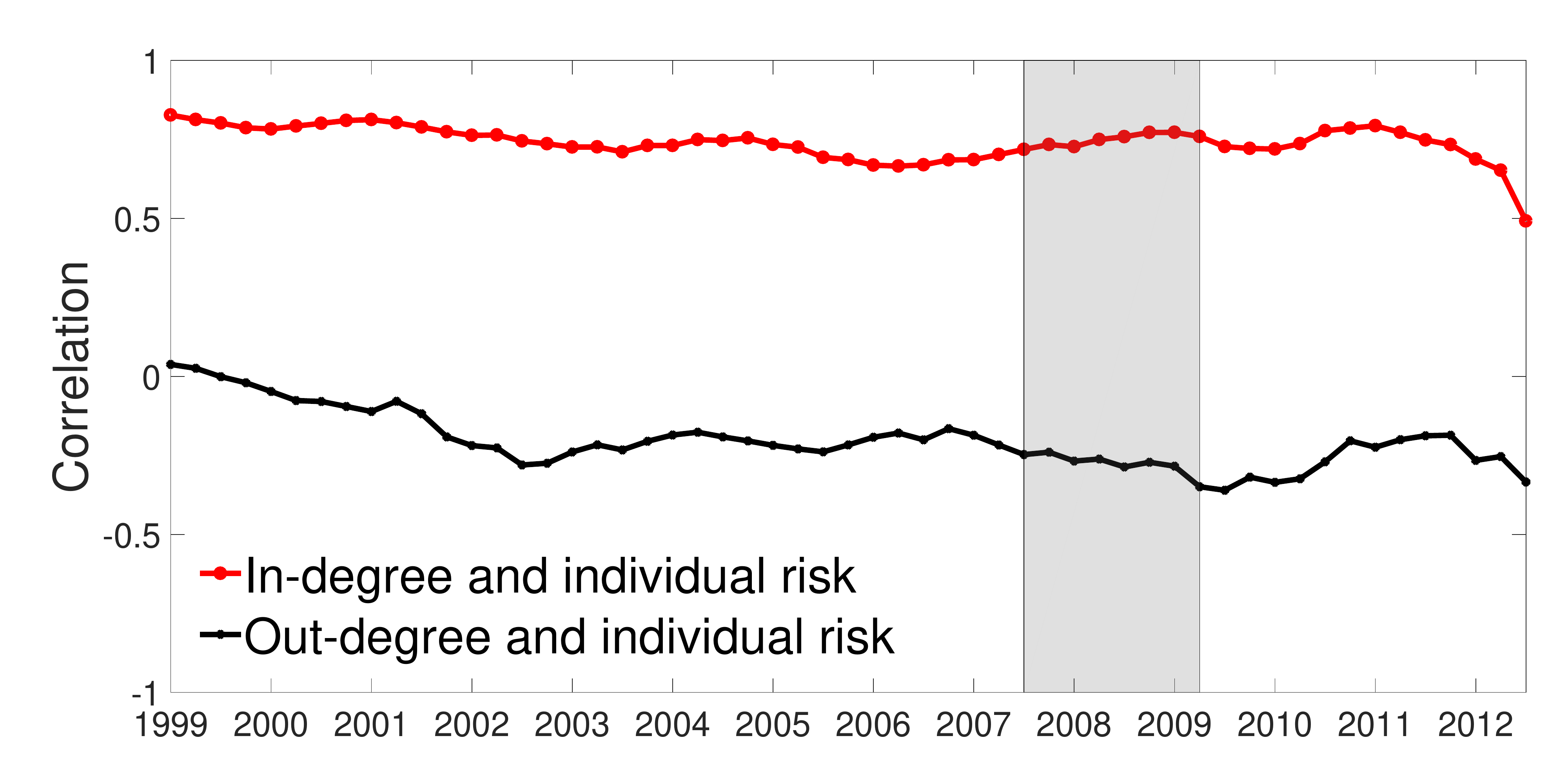}
\includegraphics[width=0.495\textwidth]{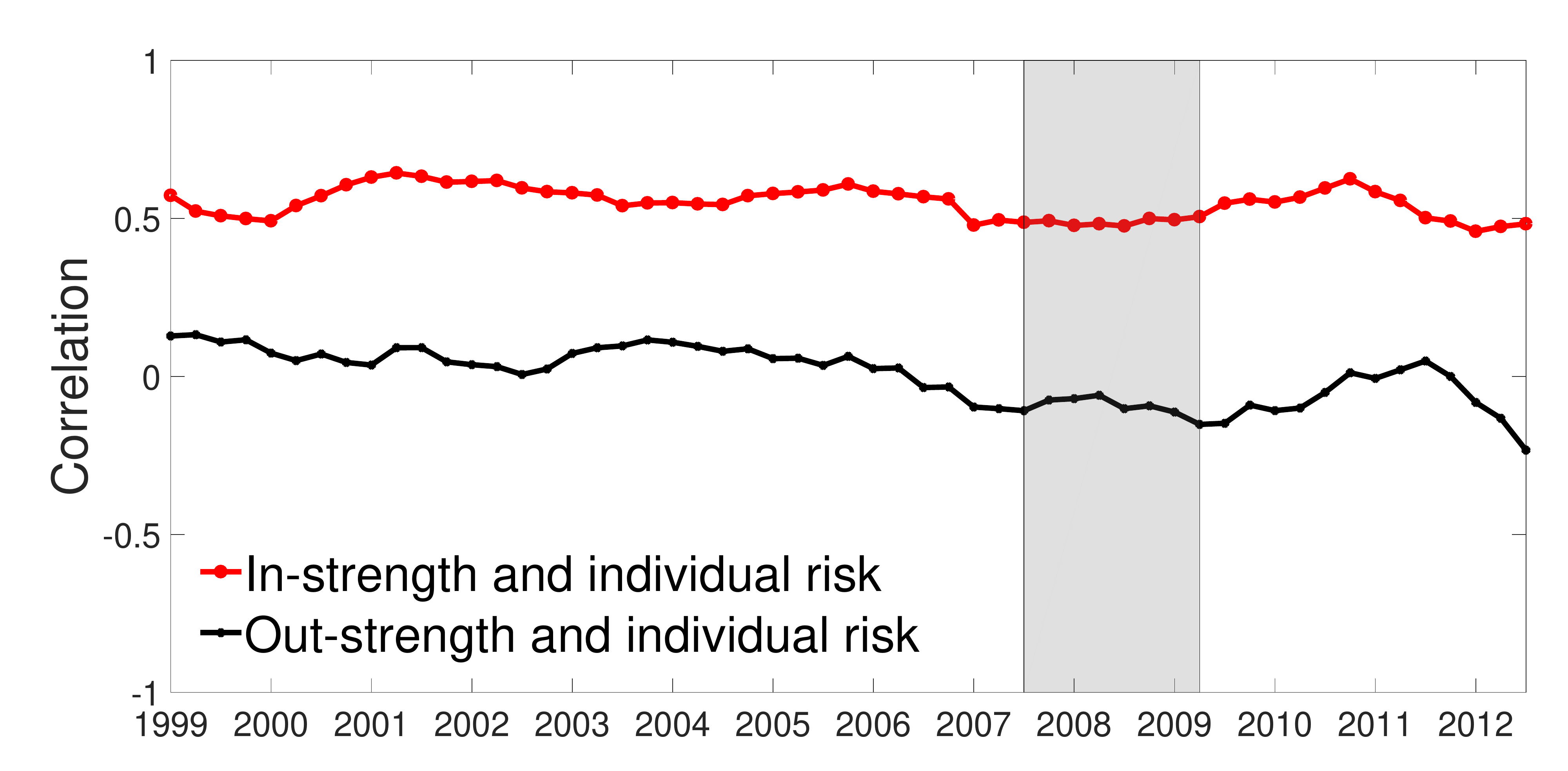}
\caption{\label{fig:7} \textbf{Time evolution of the correlation coefficients between banks network features and individual risk.} 
a) in/out-degree; b) in/out-strength. Results refer to the first economic scenario (LC-LD) and to a density of initially distressed banks equal to 1\%.}
\end{center}
\end{figure}

Besides assessing systemic risk through the number of bankrupted banks, we can measure the overall liquidity loss in the market by accounting for the size of defaulted banks. 
We thus compute the ratio between the amount of liquidity originally granted by bankrupted banks and the overall starting market liquidity, that is 
$\sum_{i\in\mathcal{B}}s_i^{O}/\sum_{i}s_{i}^{O}$.\footnote{While liquidity is in principle hoarded by both distressed and bankrupted banks, 
the stop condition for simulations is of no distressed banks left in the market (or very few of them surviving when we reach 100 steps): at that point, 
all the liquidity lost is basically provided by bankrupted banks only.} Figure \ref{fig:6} shows that the ratio of liquidity loss is remarkably high over the whole time span of the data, 
suggesting again that a small initial shock can lead to severe systemic losses. Importantly, the monetary damage is much higher than the risk metric given by the percentage of bankruptcies, 
indicating that banks accounting for larger shares of the interbank network, {\em i.e.}, banks belonging to the network core \cite{Craig2014}, are actually more likely to default.

We conclude the section by studying which network features make a bank more vulnerable to liquidity shocks. To do so we evaluate, for each quarter, the correlation coefficient 
between the in/out-degree and in-out/strength sequences of banks and their individual risk---assessed by their default frequency over the model realizations. 
Figure \ref{fig:7} shows that the liquidity riskiness of a bank is highly correlated with its in-degree (that is, the number of lenders) and in-strength (the total amount of money borrowed): 
banks that rely more on the market to fund their activities are, as expected, more vulnerable to liquidity shocks.

\section{Econometric analysis}

In this section we run a statistical exercise in order to discern the contribution of different network features of a bank on its liquidity riskiness. 
Among the various features that may affect the likelihood of bank failure, we focus on the following two set of quantities to be used as regressors in our analysis: 
\begin{itemize}
\item binary features: in-degree, out-degree, binary-clustering and coreness;
\item weighted features: in-strength, out-strength, weighted-clustering and betweenness.
\end{itemize}
In the above lists of quantities, degrees and strengths are defined in section 3.1; binary-clustering is defined in section 3.3, and weighted-clustering is obtained with the same expression 
of its binary counterpart by using $u_{ij}(t)=w_{ij}(t)+w_{ji}(t)$ \cite{Fagiolo2007}; coreness is a dummy variable taking the value of 1 if the considered bank is in the core and 0 otherwise \cite{Craig2014}; 
betweenness is the fraction of all the shortest paths in the network that pass through the considered bank \cite{Freeman1997}. 
All these quantities are measures of nodes centrality (or importance) in the network. 
Indicating as $x_i^m(t)$ the value of variable $m$ for bank $i$ at time $t$, our benchmark regression reads:
\begin{equation}\label{eq:regret}
y_{i}(t)=\omega_i+\eta(t)+\sum_m\alpha_mx_i^m(t)+\Theta_t\sum_m\beta_mx_i^m(t)+\zeta\Theta_t+\varepsilon_{i}(t),
\end{equation}
where $y_{i}(t)$, namely the default frequency in the EDB model observed for each individual bank $i$ at date $t$, is the dependent variable. 
The latter is estimated by within-group OLS estimator (Fixed Effect) with cluster-adjusted standard errors. 
In the above expression, we account for unobserved heterogeneity at bank level through the coefficient $\omega_i$, encompassing a variety of time-invariant factors such as 
bank location, institutional features of the country where it is based, ownership, and so on. On the other hand, $\eta(t)$ accounts for generic temporal effects such as aggregate shocks 
hitting the market at time $t$. We also introduce $N_t$ discontinuities through the dummy variable $\Theta_t$, taking on the value 0 before 2008Q4 (first quarter after Lehman Brothers default) 
and 1 afterwards, in order to capture any change of variables due to the global financial crisis. Overall, the sets of regression coefficients we are interested in are $\{\alpha_{m}\}$, 
indicating whether feature $m$ explains the variability of default frequencies, and $\{\beta_{m}\}$, showing how the effect of feature $m$ changes after the crisis. 

Tables \ref{tab:binary} and \ref{tab:weighted} show regression results for the set of binary and weighted variables, respectively.\footnote{We report only result for the first economic scenario LC-LD, 
since the other two scenarios produce similar results.} In both cases, we consider an initial density of distressed banks equal to $1\%$, $5\%$, $10\%$ and $20\%$ (corresponding to the different table columns). 
As a general observation, we see that bank network features, {\em i.e.}, the topology of the network, plays an increasing significant role for a decreasing initial density of distressed banks. 
This means that network effects are important especially for small initial shocks. 
We then observe that betweenness and coreness do influence the default frequency, while clustering is the only metric that does not. 
Concerning the binary features reported in Table \ref{tab:binary}, we see that the number of lenders of a given bank covariates positively with that bank's default frequency, 
whereas, the number of borrowers for that bank does it negatively. Table \ref{tab:weighted} then shows that also the total borrowing and lending are significant variables, 
however their effect is negligible compared to that of their binary counterparts. 
Finally, concerning the impact of variables after the crisis, we find a much lower statistical effect for betweenness, coreness and clustering. 
Discontinuities are instead observed for banks connectivities, with in-degree doubling its effect after the crisis. 
The impact of the crisis, represented by the variable $\Theta$, increases in both regressions with the initial density of distressed banks. 
Note however that the crisis is significant only for the binary set of regressors (and for low values of initial density): 
default frequencies drops after the crisis due to the diminished trades between banks (liquidity hoarding). 

Overall, the goodness of the two regressions in term of cluster-adjusted $R^{2}$ indicates that the binary features can explain the variability of the dependent variable 
two-three times more than the weighted features. We can thus conclude that coreness and in/out-degree are more explanatory for systemic risk than the other variables, 
in line with the recent paradigm shift from ``too big to fail'' to ``too interconnected to fail'' \cite{Huser2015}. Note that Toivanen \cite{Toivanen2013} reached similar conclusions 
for its credit-driven contagion model, in which an increase in the connectivity of a bank poses a greater threat to the banking system than an increase in its size.

{
\def\sym#1{\ifmmode^{#1}\else\(^{#1}\)\fi}
\begin {table}[p]
\caption {\textbf{Regression results for independent variables given by binary network features of banks,} 
obtained using within-group OLS estimator (Fixed Effect) with cluster-adjusted standard errors. 
Number of observations $N=4387$, number of groups $=230$, observations per group $=19$.}\label{tab:binary} 
\begin{center}
\begin{tabular}{l*{4}{c}}
\hline
&\multicolumn{4}{c}{Infection Probability}\\
            &\multicolumn{1}{c}{$1\%$}&\multicolumn{1}{c}{$5\%$}&\multicolumn{1}{c}{$10\%$}&\multicolumn{1}{c}{$20\%$}\\
\hline\hline
$\alpha$\_coreness   &       9.791\sym{***}&       8.900\sym{***}&       7.609\sym{***}&       5.993\sym{***}\\
             &     (1.695)         &     (1.610)         &     (1.494)         &     (1.298)         \\
\addlinespace
$\alpha$\_binary-clustering   &      -1.371         &      -2.412         &      -2.522         &      -2.710         \\
               &    (12.159)         &    (11.445)         &    (10.419)         &     (8.824)         \\
\addlinespace
$\alpha$\_in-degree     &       0.992\sym{***}&       0.945\sym{***}&       0.878\sym{***}&       0.752\sym{***}\\
            &     (0.084)         &     (0.084)         &     (0.081)         &     (0.073)         \\
\addlinespace
$\alpha$\_out-degree   &      -1.570\sym{***}&      -1.498\sym{***}&      -1.375\sym{***}&      -1.156\sym{***}\\
             &     (0.076)         &     (0.078)         &     (0.078)         &     (0.074)         \\
\hline
\addlinespace
$\beta$\_coreness&       -8.350\sym{**} &      -7.006\sym{*}  &      -5.388         &      -3.001         \\
           &     (3.145)         &     (3.251)         &     (3.237)         &     (3.030)         \\
\addlinespace
$\beta$\_binary-clustering &      19.172         &      17.807         &      16.411         &      14.899         \\
            &    (14.102)         &    (13.649)         &    (12.863)         &    (11.393)         \\
\addlinespace
$\beta$\_in-degree&       1.756\sym{***}&       1.761\sym{***}&       1.617\sym{***}&       1.218\sym{***}\\
           &     (0.224)         &     (0.232)         &     (0.230)         &     (0.207)         \\
\addlinespace
$\beta$\_out-degree&      -1.631\sym{***}&      -1.719\sym{***}&      -1.642\sym{***}&      -1.313\sym{***}\\
             &     (0.223)         &     (0.225)         &     (0.220)         &     (0.201)         \\
\addlinespace
\hline
$\Theta$     &     -15.702\sym{**} &     -12.779\sym{**} &      -9.293\sym{*}  &      -4.778         \\
            &     (4.883)         &     (4.793)         &     (4.564)         &     (4.116)         \\
\addlinespace
\hline
$\eta$      &      42.738\sym{***}&      49.039\sym{***}&      54.855\sym{***}&      63.231\sym{***}\\
                 &     (3.775)         &     (3.609)         &     (3.335)         &     (2.873)         \\
\midrule
\(N\)        &        4387         &        4387         &        4387         &        4387         \\
adj. \(R^{2}\)&       0.363         &       0.358         &       0.345         &       0.315         \\
\hline\hline
\multicolumn{5}{l}{\footnotesize Standard errors in parentheses}\\
\multicolumn{5}{l}{\footnotesize \sym{*} \(p<0.05\), \sym{**} \(p<0.01\), \sym{***} \(p<0.001\)}\\
\end{tabular}
\end{center}
\end{table}
}

{
\def\sym#1{\ifmmode^{#1}\else\(^{#1}\)\fi}
\begin {table}[p]
\caption {\textbf{Regression results for independent variables given by weighted network features of banks,} 
obtained using within-group OLS estimator (Fixed Effect) with cluster-adjusted standard errors. 
Number of observations $N=4387$, number of groups $=230$, observations per group $=19$.}\label{tab:weighted} 
\begin{center}
\begin{tabular}{l*{4}{c}}
\hline
&\multicolumn{4}{c}{Infection Probability}\\
            &\multicolumn{1}{c}{$1\%$}&\multicolumn{1}{c}{$5\%$}&\multicolumn{1}{c}{$10\%$}&\multicolumn{1}{c}{$20\%$}\\\\
\hline\hline
$\alpha$\_betweenness&       0.045\sym{***}&       0.044\sym{***}&       0.041\sym{***}&       0.036\sym{***}\\
           &     (0.007)         &     (0.007)         &     (0.006)         &     (0.006)         \\
\addlinespace
$\alpha$\_weighted-clustering&       0.016         &       0.015         &       0.014         &       0.012         \\
                &     (0.024)         &     (0.021)         &     (0.018)         &     (0.013)         \\
\addlinespace
$\alpha$\_in-strength   &       0.003\sym{***}&       0.003\sym{***}&       0.002\sym{***}&       0.002\sym{***}\\
           &     (0.001)         &     (0.001)         &     (0.000)         &     (0.000)         \\
\addlinespace
$\alpha$\_out-strength   &      -0.003\sym{***}&      -0.003\sym{***}&      -0.002\sym{***}&      -0.002\sym{***}\\
           &     (0.000)         &     (0.000)         &     (0.000)         &     (0.000)         \\
\addlinespace
\hline
$\beta$\_betweenness&       0.042\sym{**} &       0.045\sym{**} &       0.043\sym{**} &       0.036\sym{*}  \\
            &     (0.015)         &     (0.015)         &     (0.016)         &     (0.015)         \\
\addlinespace
$\beta$\_weighted-clustering&       0.066         &       0.078         &       0.075         &       0.049         \\
            &     (0.116)         &     (0.112)         &     (0.102)         &     (0.084)         \\
\addlinespace
$\beta$\_in-strength&       0.000         &       0.001         &       0.001         &       0.000         \\
            &     (0.001)         &     (0.001)         &     (0.001)         &     (0.001)         \\
\addlinespace
$\beta$\_out-strength&      -0.004\sym{***}&      -0.004\sym{***}&      -0.004\sym{**} &      -0.003\sym{**} \\
            &     (0.001)         &     (0.001)         &     (0.001)         &     (0.001)         \\
\hline
\addlinespace
$\Theta$      &      -0.763         &       1.145         &       2.909         &       4.569\sym{*}  \\
           &     (2.533)         &     (2.556)         &     (2.473)         &     (2.202)         \\
\hline
\addlinespace
$\eta$    &      45.784\sym{***}&      51.676\sym{***}&      57.537\sym{***}&      65.698\sym{***}\\
            &     (1.375)         &     (1.312)         &     (1.205)         &     (1.016)         \\
\midrule
\(N\)       &        4387         &        4387         &        4387         &        4387         \\
adj. \(R^{2}\)&       0.168         &       0.162         &       0.153         &       0.139         \\
\hline\hline
\multicolumn{5}{l}{\footnotesize Standard errors in parentheses}\\
\multicolumn{5}{l}{\footnotesize \sym{*} \(p<0.05\), \sym{**} \(p<0.01\), \sym{***} \(p<0.001\)}\\
\end{tabular}
\end{center}
\end{table}
}

\section{Conclusions}

In this work we have addressed the issue of systemic risk in interbank markets due to liquidity shortcomings. 
Using transaction data from the Italian electronic market for interbank deposits (e-MID), we have analyzed the interbank network structure, 
and studied its resilience to liquidity shocks through the introduction of the {\em Exposed-Distressed- Bankrupted} (EDB) model---simulating an epidemic dynamics of liquidity disease within the market. 
We found that the 2007/2008 global financial crisis had a significant impact on the network topology: liquidity hoarding policies and worries of counterparty creditworthiness 
resulted not only in a poorer interbank market in terms of active participants and density/extent of transactions, but also in structural changes of the market itself 
such as the disappearance of hub banks and preferential lending channels---which ultimately led the interbank network to become less efficient in terms of monetary flow between banks. 
Indeed, results of the EDB model show that systemic risk was maximal at the onset of the global financial crisis, and the subsequent financial burst forced banks to change their strategic behavior 
to adapt to the new harsher environment, ultimately lowering risk but depriving the market from its functionalities. 
We also point out that dynamics of liquidity shocks reverberation on the interbank network are extremely important to consider in stress-test scenarios, 
as the overall losses in the system can be severe also for weak initial shocks. The failure probability turns out to be enhanced for banks 
that rely heavily on the market for liquidity provision, especially in terms of number of borrowers: a high diversification of funding channels 
usually minimize roll-over risk for low level of financial distress, but during crisis periods it increases the probability to be hit by a liquidity shock.

The EDB model we implement in this work can be further generalized by acting on the infection and default probabilities, 
which can be modified to include exogenous spillovers. For instance, the default rate could be made dependent on depositors withdraws, 
or include bailout mechanisms that bring a bank back to the exposed state. Finally note that, as we already stated, the EDB model is implemented without considering the detailed 
balance sheet positions of banks, to which we do not have access through e-MID data. While such a simplification is not crucial when studying qualitatively 
the financial (in)stability of the market in relation with its network topological structure, the model is open to account for banks balance sheet information, 
in order to properly assess the real extent of liquidity-driven systemic risk in interbank markets. 

\section*{Acknowledgments}

We thank Alessandro Belmonte for useful discussions. Giulio Cimini and Riccardo Di Clemente acknowledge support from project GROWTHCOM (FP7-ICT, n. 611272). 
Giulio Cimini also acknowledges support from projects MULTIPLEX (FP7-ICT, n. 317532) and DOLFINS (H2020-EU.1.2.2., n. 640772). 
The funders had no role in study design, data collection and analysis, decision to publish, or preparation of the manuscript.


\end{document}